\documentclass[a4paper,usenatbib,usegraphicx,useAMS,fleqn]{mn2e}
\usepackage[backref,breaklinks]{hyperref}

\usepackage{amsfonts,amsmath,amssymb,gensymb}
\usepackage{natbib}
\usepackage{aas_macros,apjfonts}
\usepackage{rotating,array,multirow,array,longtable,tabularx,color,bm}
\usepackage{pstricks,epsf, epsfig}
\usepackage{afterpage,fix2col}
\newcommand{\msun}{M_{\odot}}

\author[Gerosa \& Sesana]{ 
Davide Gerosa$^{1} \thanks{E-mail: d.gerosa@damtp.cam.ac.uk}$ 
and
Alberto Sesana$^{2} \thanks{E-mail: alberto.sesana@aei.mpg.de}$ \\
$^1$Department of Applied Mathematics and Theoretical Physics, Centre for Mathematical Sciences, \\ $\:\;$University of Cambridge, Wilberforce Road, Cambridge CB3 0WA, UK\\
$^2$Max-Planck-Institut f{\"u}r Gravitationsphysik, Albert Einstein Institut, Am M{\"u}lenber
1, 14476 Golm, Germany
}

\title[{Missing black holes in BCGs}]{
{Missing black holes in brightest cluster galaxies as evidence for the occurrence of superkicks in nature}
}

\begin{document}

\maketitle 

\begin{abstract}
{ We investigate the consequences of superkicks on the population of supermassive black holes (SMBHs) in the Universe residing in brightest cluster galaxies (BCGs). There is strong observational evidence that BCGs grew prominently at late times (up to a factor 2-4 in mass from $z=1$), mainly through mergers with satellite galaxies from the cluster, and they are known to host the most massive SMBHs ever observed. Those SMBHs are also expected to grow hierarchically, experiencing a series of mergers with other SMBHs brought in by merging satellites. Because of the net linear momentum taken away from the asymmetric gravitational wave emission, the remnant SMBH experiences a kick in the opposite direction. Kicks may be as large as ~5000 Km s$^{-1}$ (``superkicks''), pushing the SMBHs out in the cluster outskirts for a time comparable to galaxy-evolution timescales. We predict, under a number of plausible assumptions, that superkicks can efficiently eject SMBHs from BCGs, bringing their occupation fraction down to a likely range $0.9<f<0.99$ in the local Universe. Future thirty-meter-class telescopes like ELT and TMT will be capable of measuring SMBHs in hundreds of BCGs up to $z=0.2$, testing the occurrence of superkicks in nature and the strong-gravity regime of SMBH mergers.}

\end{abstract}

\begin{keywords}
Black hole physics - galaxies: evolution - galaxies: interactions - gravitational waves.
\end{keywords}

\section{Introduction}
\label{sec_intro}

The centers of galaxy clusters host the most massive galaxies in the Universe, generally known as brightest cluster galaxies (BCGs) 
Their luminosity can easily exceed $10^{12} L_\odot$ and, consequently, their estimated masses can be up to few$\times10^{12}\msun$. They also host the biggest supermassive black holes (SMBHs) known in the Universe, with masses in the range $10^9-10^{10}\msun$ \citep{2012ApJ...756..179M}, tipping the observed SMBH-host relations  at the high mass end \citep{2013ApJ...764..184M}. 

In the context of the $\Lambda$ cold dark matter ($\Lambda$CDM) cosmological paradigm, large dark matter (DM) halos in the Universe build up hierarchically \citep{1978MNRAS.183..341W}, driving the assembly of galactic structures. Galaxy formation kicks off at high redshifts, as gas starts to cool at the centers of DM halos. Following the halo hierarchy, small protogalaxies merge with each other forming larger ones. This process continues until the present time, resulting in the formation of massive galaxies we see today. Within this framework, also SMBH grow hierarchically, experiencing a sequence of accretion events and merging with other SMBHs following galaxy mergers \citep{1980Natur.287..307B,2003ApJ...582..559V}.

One interesting astrophysical consequence of SMBH binary mergers is the gravitational recoil. Emission of asymmetric gravitational waves (GWs) in the late inspiral and final coalescence takes away net linear momentum from the binary system, and the remnant SMBH is consequently kicked in the opposite direction. With the advent of numerical relativity \citep{2005PhRvL..95l1101P,2006PhRvL..96k1101C,2006PhRvL..96k1102B}, it is now possible to simulate SMBH mergers in full general relativity and assess the magnitude of these kicks. Surprisingly, configurations have been found in which the final kick can reach magnitudes up to $\sim5000$km/s \citep{2007ApJ...659L...5C,2007PhRvL..98i1101G,2011PhRvL.107w1102L}{\footnote{{Technically, \cite{2007ApJ...659L...5C,2007PhRvL..98i1101G} found recoils up to $\sim4000$km/s for systems with spins lying in the binary orbital plane, which they referred to as "superkicks". "Hangup kicks" up to $\sim5000$km/s were found by \cite{2011PhRvL.107w1102L} in a different configuration, in which the spins are inclined with respect to the orbital plane of the binary. For simplicity, we will generally refer to high-velocity recoils as "superkicks" throughout the paper.}}} opening the possibility of SMBH ejection even from the deepest potential wells created by the most massive galaxies \citep{2004ApJ...607L...9M,2007ApJ...662L..63S}. { Observationally, few candidate recoiling SMBHs have been recently identified as off-center AGNs \citep{2010ApJ...717..209C,2012ApJ...752...49C,2014arXiv1401.6798K}, and an excellent review of the spatial and kinematical observational signatures of these peculiar systems can be found in \cite{2012AdAst2012E..14K}. 
A direct consequence of high velocity kicks is that} the SMBH occupation fraction may be altered \citep{2007ApJ...667L.133S,2008MNRAS.384.1387V,2010MNRAS.404.2143V}, providing an indirect way to test the strong-gravity physics behind GW kicks. In this paper, we explore this possibility by investigating the consequences of gravitational recoils onto  SMBH masses and the occupation fraction in BCGs.

Although kicks will naturally eject SMBHs more easily from lighter galaxies \citep[as extensively investigated by][]{2010MNRAS.404.2143V}, there are at least three good reasons for considering this possibility in BCGs. Firstly, BCGs show the strongest mass evolution from $z\approx 1.5$ up to now. In general, both detailed numerical simulations of galaxy formation \citep{2007MNRAS.375....2D,2010ApJ...725.2312O,2012MNRAS.425..641L} and observations of BCGs at different $z$ \citep{2011MNRAS.415.3903T,2012MNRAS.427..550L,2013MNRAS.433..825L}, show an average mass doubling from $z=1$ to the present time. Though it is difficult to assess observationally what is the cause of this mass growth, it appears in simulations to be driven primarily by galaxy mergers \citep{2011ApJ...742..103L,2013MNRAS.435..901L}. This is also consistent with close galaxy pair counts at $z<1$ \citep{2006ApJ...652..270B,2009ApJ...697.1369B,2009A&A...498..379D,2010ApJ...719..844R,2012ApJ...747...85X,2012A&A...548A...7L}, which imply a prominent merger activity for these systems. In contrast with all other types of galaxies, very massive ellipticals (and BCGs in particular) are expected to have undergone several mergers in the last 10Gyr, some of which 'major' (i.e. with satellite to primary galaxy mass ratio $M_2/M_1>1/4$). It is therefore possible that they also experienced a few SMBH binary coalescences, with consequent gravitational recoils. Secondly, SMBHs of mass $>10^{9}\msun$ in the relatively low-density environment of BCG nuclei have the largest impact on the dynamics of the surrounding stars \citep{2012ApJ...756..179M}. The influence radius of the SMBH can be up to few hundred parsecs, making them ideal targets for direct dynamical measurements of SMBH masses. With angular resolutions of $\approx0.1$arcsec, it is today possible to confidently measure SMBH masses in BCGs up to $z\approx0.03$. A factor of ten improvement in the instrumentation, expected with the Thirty Meter Telescope (TMT) and the European Extremely Large Telescope (ELT), will dramatically increase this range. As an example, \cite{2014AJ....147...93D} estimated that 50 masses of SMBHs residing in BCGs up to $z=0.05$ can be measured with a relatively cheap program of 14 observing nights on the TMT. Moreover, they show that the TMT potential will be much greater than that, making mass measurement possible in hundreds of BCGs up to $z\approx0.2$. Conversely, in Milky Way-type galaxies with SMBH sphere of influence of the order of few parsecs, even with ELT precision dynamical measurements will be restricted to our local neighborhood ($D<30$ Mpc, $z<0.01$). Lastly, according to our galaxy formation knowledge, the SMBH occupation fraction $f$ (i.e., the fraction of galaxies hosting a SMBH) is an increasing function of the galaxy mass. Although already at dwarf galaxy scales $f$ might be around unity \citep{2011ApJ...742...13B}, observations of galaxies in Virgo galaxies shows a sudden drop in the  X-ray activity at stellar masses around $10^{10}\msun$ \citep{2012ApJ...747...57M}. Although this cannot be taken as evidence of lack of nuclear SMBHs, there is no observational confirmation of a large $f$ for galaxies on those small scales.

{Some tentative candidates of SMBH ejections from BCGs have already been identified: the BCG in the A2261 cluster  shows an exceptionally large core of $3.2$~kpc consistent  with the absence of a scouring SMBH  \citep{2012ApJ...756..159P}; the small $1.2 \times 10^{11} \msun$ lenticular galaxy NCG 1277 in the Perseus cluster hosts an exceptionally  heavy SMBH of $1.7 \times 10^{10} \msun$ \citep{2012Natur.491..729V} which may have been grown in the close BCG NCG 1275, ejected by a superkick and finally captured by NCG 1277 \citep{2013ApJ...772L...5S}.}

Summarizing, BCGs, being the most massive galaxies in the Universe, 
(i) are expected to have $f=1$; 
(ii) have possibly experienced multiple mergers at low redshift; 
(iii) are the easiest targets for nuclear SMBH mass measurements. 
These facts make them ideal targets for observing the effects of extreme recoils: any observational confirmation of a missing nuclear SMBH would provide strong evidence for the occurrence of superkicks. 

The paper is organized as follows.  Sec.~\ref{sec_model} presents the ingredients of our models:  (i) SMBH merger fitting formulas; (ii) galaxy density profiles; (iii) prescriptions for the SMBH return timescales and  (iv) the merger events; and (v) finally our evolutionary procedure.  We highlight our results in Sec.~\ref{sec_results}  and present our conclusions  in Sec.~\ref{sec_concl}
Throughout this paper, we use a $\Lambda$CDM cosmological model with $\Omega_{M}=0.27$, $\Omega_\Lambda=0.73$ and $H_0=100 h {\rm \,km/s\,Mpc}^{-1}=70 {\rm \,km/s\,Mpc}^{-1}$.

%%%%%%%%%%%%%%%%%%%%%%%%%%%%%%%%%%%%%%%%%%
\section{Brightest-cluster-galaxy merger modeling}
\label{sec_model}

A thoughtful modeling of the recoil effect on the SMBH occupation fraction in BCGs requires to put together in a coherent framework four main ingredients:
\begin{itemize}
\item the recoil magnitude as a function of the SMBH binary parameters (binary mass ratio, magnitude and orientation of the individual SMBH spins);
\item the gravitational potential in which the recoiled SMBH evolves;
\item the return timescale for SMBHs suffering kicks below the escape velocity of their hosts;
\item the number of mergers experienced by BCGs as a function of $z$ and of the galaxy mass ratio.
\end{itemize}
We will describe each item separately in the following subsections, providing in Sec.~\ref{sec2.5} a description of the `coherent framework' that brings them together; we point the readers not interested in all the mathematical details of our model directly to that section.

%%%%%%%%%%%%%%%%%%%%%%%%%%%%%%%%%%%%%%%%%%
\subsection{Black-hole final mass, spin and kick velocity}
\label{bhmergermod}
We start with modelling the properties of the remnant SMBH as a function of the properties of the progenitor merging holes.  We use a standard notation in which $m_1$ and $m_2$ denote the individual masses  of the merging SMBHs (with $m_1>m_2$), $M=m_1+m_2$ is the total mass, $q=m_2/m_1\leq1$ is the mass ratio and $\eta=m_1 m_2/M^2$ is the symmetric mass ratio.  The SMBH spin vectors are (with $i=1,2$)
\begin{align}
\mathbf{S_i}= \chi_i \frac{G m_i^2}{c}\mathbf{\hat S_i},
\end{align}
where $0\leq\chi_i\leq1$ is the  dimensionless-spin parameter  and hats denote unit vectors. We describe the directions of the spins $\mathbf{\hat S_i}$ with three angles $\theta_1,\theta_2$ and $\Delta\Phi$ defined to be (cf. Fig.~1 in \citealt{2014PhRvD..89l4025G})
\begin{align}
\begin{aligned}
\cos\theta_1=\mathbf{\hat S_1} \cdot \mathbf{\hat L}\,,
\quad
 \cos\theta_2=\mathbf{\hat S_2} \cdot \mathbf{\hat L}\,,
\quad
\cos\Delta\Phi=\frac{\mathbf{\hat S_1} \times \mathbf{\hat L}}{|\mathbf{\hat S_1} \times  \mathbf{\hat L} |} \cdot 
\frac{\mathbf{\hat S_2} \times \mathbf{\hat L}}{|\mathbf{\hat S_2} \times \mathbf{\hat L} |}\,,
\end{aligned}
\end{align}
where $\mathbf{\hat L}$ is the (instantaneous) direction of the orbital angular momentum of the binary. It is also useful to define the following quantities
\begin{align}
\mathbf{\Delta}&=\frac{q \chi_2 \mathbf{\hat S_2} -  \chi_1 \mathbf{\hat S_1}}{1+q}\,,
\qquad
\mathbf{\tilde{\boldsymbol{\chi}}}=\frac{q^2 \chi_2 \mathbf{\hat S_2} + \chi_1 \mathbf{\hat S_1}}{(1+q)^2},
\label{chitilde}
\end{align}
and to introduce the subscripts $\parallel$ and $\perp$ for vector components along/perpendicular to the orbital angular momentum of the binary:  
\mbox{$\tilde\chi_\parallel = \mathbf{\tilde{\boldsymbol{\chi}}}\cdot\mathbf{\hat L} $},
\mbox{$\tilde\chi_\perp = | \mathbf{\tilde{\boldsymbol{\chi}}}\times \mathbf{\hat L} |$},
\mbox{$\Delta_\parallel = \mathbf{\Delta}\cdot\mathbf{\hat L} $},
\mbox{$\Delta_\perp = | \mathbf{\Delta}\times \mathbf{\hat L} |$}.

The energy radiated during the inspiral and merger phase $E_{\rm rad}$ reduces the post-merger mass to $M_f=M-E_{\rm rad}c^{-2}$. The dependence of $E_{\rm rad}$ on the initial parameters (namely the masses and the spins) can be derived analytically in the test-particle limit $q\rightarrow 0$ \citep{2008PhRvD..78h4030K}, while the comparable-mass regime $q \simeq 1$ can only be estimated using full numerical relativity simulations \citep{2007PhRvD..76f4034B,2008PhRvD..78h1501T,2010CQGra..27k4006L}. Here we use the expression recently provided by \cite{2012ApJ...758...63B}, in which the two regimes are interpolated
\begin{align}
\frac{E_{\rm rad}}{M}&=1-\frac{M_f}{M}=\eta \left[ 1- E'_{\rm ISCO}\right] \notag
\\
& +4\eta^{2} \left[4 p_0 +16 p_1 \tilde \chi_\parallel \left(\tilde \chi_\parallel +1\right) +E'_{\rm ISCO} -1 \right] \label{erad}\,,
\end{align}
where $c^2 E'_{\textsc{isco}}$ is the energy per unit mass at the innermost stable circular orbit (ISCO) in the test-particle limit generalized to inclined orbits and evaluated at the effective spin $\mathbf{\tilde{\boldsymbol{\chi}}}$  
\citep{1973blho.conf..215B}
\begin{align}
E'_{\textsc{isco}} &= \sqrt{1-\frac{2}{3 r'_{\textsc{isco}}}}\,, \\
r'_{\textsc{isco}}&=3+ Z_2
- {\rm sign}(\tilde \chi_\parallel)\sqrt{(3-Z_1)(3+Z_1+2Z_2)}\,,
\\
Z_1&=1+\left(1-\tilde \chi_\parallel^2\right)^{1/3}\left[\left(1+\tilde \chi_\parallel\right)^{1/3} + \left(1-\tilde \chi_\parallel\right)^{1/3}\right]\,,
\\
Z_2&=\sqrt{3\tilde \chi_\parallel^2 + Z_1^2}.
\end{align}
The parameters $p_0$ and $p_1$ in Eq.~(\ref{erad}) were fitted by \cite{2012ApJ...758...63B} using  the numerical relativity data published at the time (see references therein): they report $p_0=0.04827$ and $p_1=0.01707$.

The final spin magnitude $\chi_f$ has been predicted either by calibrating fitting formulas with numerical relativity simulations \citep{2008PhRvD..78h1501T,2008PhRvD..78d4002R,2009ApJ...704L..40B,2010CQGra..27k4006L}, or by extrapolating test-particle results \citep{2008PhRvD..77b6004B,2008PhRvD..78h4030K}. Here we use the expression developed by \cite{2009ApJ...704L..40B}, which has been shown to reproduce the available numerical relativity data with $8\%$ precision in $\chi_f$ for every value of $q$:
\begin{align}
\chi_f&= \left|\mathbf{\tilde{\boldsymbol{\chi}}} + \frac{q}{(1+q)^2}\ell\, \mathbf{\hat L}\right| \label{finalspin}\,,
\\
\ell &= 2\sqrt{3} + t_2 \eta + t_3 \eta^2 + s_4 \frac{(1+q)^4}{(1+q^2)^2} \tilde \chi^2
+ (s_5 \eta + t_0 +2)\frac{(1+q)^2}{1+q^2} \tilde \chi_\parallel\,.
\end{align}
The remaining free parameters are fitted to numerical relativity simulations (see \citealt{2009ApJ...704L..40B} for details): $t_0=-2.8904$, $t_2=-3.51712$, $t_3=2.5763$, $s_4=-0.1229$ and $s_5=0.4537$.
We assume $\chi_f=1$ whenever the fitting formula  (\ref{finalspin}) predicts higher unphysical values.

GW recoils generally arise from asymmetries in the merging binary, that could be either in the masses or in the spins. Fitting formulas for the recoil velocity $\mathbf{v_k}$ are typically broken down into a mass asymmetry term $v_m$, and two spin asymmetry terms $v_{s\parallel}$ and $v_{s\perp}$ \citep{2007ApJ...659L...5C}
\begin{align}
\mathbf{v_k}=v_m \mathbf{\hat e_{\perp 1}} +v_{s\perp} (\cos\xi \mathbf{\hat e_{\perp 1}} +\sin\xi \mathbf{\hat e_{\perp 2}}) + v_{s\parallel} \mathbf{\hat L} \label{kickvel}\,,
\end{align}
where $\mathbf{\hat e_{\perp 1}},\mathbf{\hat e_{\perp 2}}$ are two orthogonal unit vectors in the orbital plane and $\xi$ is the angle between the mass term and the orbital-plane spin term.  Expressions for $v_m,v_{s\parallel}$ and $v_{s\perp}$ are available as fitting formulas to the numerical simulations. In this work we implement the following expressions  
\begin{align}
v_m&= A \eta^2\frac{1-q}{1+q}(1+ B\eta)\,,  \\
v_{s\perp}&= H \eta^2 \Delta_\parallel\,, \\
v_{s\parallel}&=16 \eta^2  [ 
\Delta_\perp (V_{11} + 2 V_{ A} {\tilde \chi_\parallel}+4 V_{ B} {\tilde \chi_\parallel^2}+8 V_{ C} {\tilde \chi_\parallel^3}) \notag \\
&+{\tilde \chi_{\perp}}\Delta_\parallel  (2 C_2 + 4 C_3 {\tilde \chi_{\parallel}})
] \cos{\Theta} \,.
\label{vparallel}
\end{align}
The term proportional to $V_{11}$ in Eq.~(\ref{vparallel}) arises from the superkick formula  \citep{2007PhRvL..98w1101G,2007ApJ...659L...5C}, the terms in $V_{A,B,C}$ have been called ``hangup-kick'' effect \citep{2011PhRvL.107w1102L}, while the ones proportional to $C_{2,3}$ model the newly discovered ``cross-kick'' effect \citep{2013PhRvD..87h4027L}. The parameters in the equations above are currently estimated to be: $A = 1.2 \times 10^4~ {\rm km/s}$, $B = -0.93$ \citep{2007PhRvL..98i1101G}, $H=6.9  \times 10^3 ~{\rm km/s}$ \citep{2008PhRvD..77d4028L},  $V_{11} = 3677.76 ~{\rm km/s}$, $V_A  = 2481.21~ {\rm km/s}$, $V_B  = 1792.45 ~{\rm km/s}$, $V_C  = 1506.52 ~{\rm km/s}$ \citep{2012PhRvD..85h4015L}, $C_2=1140 ~{\rm km/s}$, $C_3=2481~ {\rm km/s}$ \citep{2013PhRvD..87h4027L}, $\xi=145^\circ$ \citep{2008PhRvD..77d4028L}. The value of the angle $\Theta$ actually depends on the initial separation of the binary in the numerical simulations: as in previous studies \citep{2012PhRvD..85h4015L,2012PhRvD..85l4049B}, we deal with this dependence by sampling over a uniform distribution in $\Theta$.

Since the spin angles $\theta_1, \theta_2$ and $\Delta\Phi$ evolve during the inspiral, the recoil fitting formula provided above can only by applied close to merger, at separations $a\sim {10} M$ where numerical relativity simulations typically start\footnote{The effect of PN resonances is critical to compute the kick velocity, but not so critical in the case of the final mass -- Eq. (\ref{erad}) -- and the final spin -- Eq. (\ref{finalspin}) --: see \cite{2009ApJ...704L..40B} for a discussion of this point. 
}. \cite{2010ApJ...715.1006K} pointed out that substantial recoil suppression/enhancement could occur due to spin-orbit resonances \citep{2004PhRvD..70l4020S} in the post-Newtonian (PN) regime of the inspiral. Spin-orbit resonances mostly affect binaries with asymmetric spin directions at large separation ($\theta_1 \neq \theta_2$), while symmetric configurations ($\theta_1 \simeq \theta_2$) are generally unaffected \citep{2013PhRvD..87j4028G}. Both effects are generally present for isotropic distributions of the spin angles, that are therefore maintained qualitatively isotropic by the PN evolution \citep{2007ApJ...661L.147B,2010PhRvD..81h4054K}.
Resonant effects are therefore strongly dependent on early-time alignment processes, such those arising from accretion-disk interactions \citep{2009MNRAS.399.2249P,2010MNRAS.402..682D,2013MNRAS.429L..30L,2013ApJ...774...43M}. 

In the present astrophysical application to BCG galaxies, we  assume isotropic distributions of both the spin vectors, taking the spin angles uniformly distributed in $\cos\theta_1,\cos\theta_2$ and $\Delta\Phi$. This is a delicate point because the misalignment distribution (also needed to properly initialize the late-time PN inspiral) has a strong impact on the recoil velocities. Although spin alignment is expected to occur when a SMBH binary is surrounded by a cold massive circumbinary disk, the relative cold gas content of galaxies is a decreasing function of their mass \citep{2010MNRAS.403..683C} and BCGs are extremely gas-poor systems. Fresh cold gas can be naturally brought in by the merging satellite; however, most of the companions of massive elliptical galaxies in observed galaxy pairs are red (up to about 70\%, \citealt{2012A&A...548A...7L}), making dry mergers the more common mass growth channel for BCGs. {Nonetheless, a fraction of mergers can still result in significant accretion onto the central SMBH; in fact, BCGs are known to power luminous radio jets \citep{2007MNRAS.379..894B} creating X-ray cavities in a number of clusters \citep{2013MNRAS.432..530R,2013MNRAS.431.1638H}. However, as a result of the `anti-hierarchical' behavior of AGNs, only about one in a thousand of the SMBHs with $M>3\times10^8\msun$ is accreting at more than 1\% of the Eddington rate at low redshift \citep{2004ApJ...613..109H}. This is despite the fact very massive galaxies experience (as we will see below) a prominent merger activity at $z<1$. Assuming one merger per BCG since $z=1$, the numbers above imply that BCGs are, on average, accreting at about 1\% of the Eddington rate for $\sim 10^{7}$ yr, resulting in a mass growth $<1\%$. This is generally insufficient to align the spins of a putative SMBH binary even if the gas is accreted by a coherent circumbinary pool as envisaged by \cite{2010MNRAS.402..682D}. Moreover, accretion might occur in a series of subsequent episodes with incoherent angular momenta orientations \citep{2006MNRAS.373L..90K,2014arXiv1402.7088S}, and disk spin alignment might be less effective than generally assumed in simple $\alpha$-disks models \citep{2013MNRAS.429L..30L}. Therefore, disk-driven alignment processes should be less important for the systems relevant to our investigation, and random spin orientation is a sensible working hypothesis for the majority of them.} In this case, the kick distribution is only weakly modified by the PN inspiral (cf. \citealt{2012PhRvD..85l4049B}, their Fig.~2) and can therefore be neglected. We checked and confirm this conclusion using the numerical PN code presented by  \cite{2013PhRvD..87j4028G}. This is particularly important because following the full PN evolution is computational expensive; by-passing this stage allows us to simulate a larger number of galaxies, thus reducing the statistical error on the final occupation fractions. For reasonably large samples ($\sim 1000$ BCGs), uncertainties in the occupation fraction  are still dominated by Poisson counting errors, rather than the PN influence on the kicks.

%%%%%%%%%%%%%%%%%%%%%%%%%%%%%%%%%%%%%%%%%%
\subsection{BCG mass-density and potential profile}
\label{bcgmodel}
BCGs sit at the center of their host cluster. The relevant potential is therefore given by the spheroidal component of the BCG plus the whole cluster DM halo.

\begin{figure}
\includegraphics[width=\columnwidth]{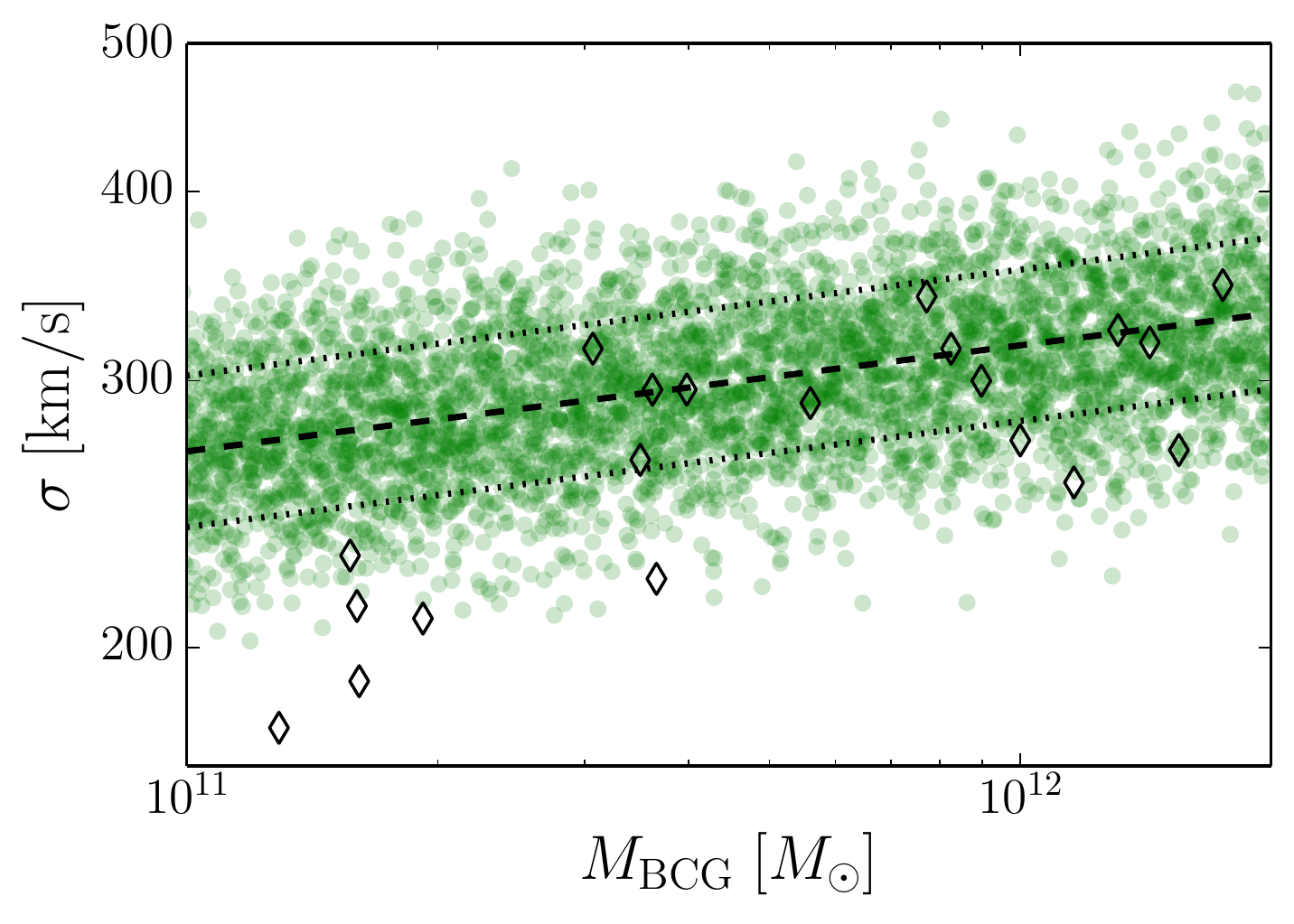}
\caption{(color online) BCG kinematical properties, modeled using the Hernquist profile. The velocity dispersion values predicted from our model  are compared with the sample of observations reported by 
\protect\cite{2013ApJ...764..184M}  
(black diamonds). Green circle points are computed sampling Eq.~(\ref{breakradius}) with a Gaussian error of $0.1$ dex and then considering $\sigma\approx0.3\sqrt{GM_{\rm BCG}/r_H}$ \protect\citep{1990ApJ...356..359H}; black dashed and dotted lines show the average and the 1-$\sigma$ interval of the same distribution.} 
\label{her_validation}
\end{figure}

A simple analytic model to describe the spheroidal component is given by the Hernquist mass-density profile (\citealt{1990ApJ...356..359H,1994AJ....107..634T},  see  \citealt{2013MNRAS.435..901L} for a specific application to BCGs)
\begin{align}
\rho_{\rm BCG}(r)&=\frac{M_{\rm BCG}}{2\pi}\frac{r_H}{r}\frac{1}{(r+r_H)^3},
\label{rhobcg}
\end{align}
where $M_{\rm BCG}$ is the mass of the spheroid and $r_H$ is a scale radius.  The scale radius $r_H$ can be related to the typical cusp radius $r_\gamma$ observed in the luminosity profiles of elliptical galaxies \citep{1997ApJ...481..710C,2007ApJ...662..808L}.
We match cusp-radius measurements from \cite{2007ApJ...662..808L} and galaxy-mass measurements from \cite{2013ApJ...764..184M}, obtaining a final sample of 14 BCGs. We fit these values using a log-log relation, obtaining 
\begin{equation}
{\rm log}\left(\frac{r_\gamma}{\text{pc}}\right)=-7.73+0.857{\rm log}\left(\frac{M_{\rm BCG}}{M_\odot}\right),
\label{breakradius}
\end{equation}
with dispersion of 0.1 dex. The central densities  of elliptical cores typically lie in the range $10^3-10^4\msun$/pc$^3$ \citep[see, e.g.,][]{2005MNRAS.362..197T};
 these values are reproduced by scaling the cusp radius by an order of magnitude, i.e. taking $r_H=10 r_\gamma$. This choice gives acceptable results in terms of the kinematical properties of BCGs, especially at typical BCG masses $\sim 10^{12} M_\odot$: Fig.~\ref{her_validation} shows the velocity dispersion of the BCG $\sigma\approx0.3\sqrt{GM_{\rm BCG}/r_H}$ 
  \citep{1990ApJ...356..359H} compared\footnote{{Since the baryonic structure is much more concentrated that the DM halo (i.e. $r_H\ll r_v$), considering the stellar component only is sufficient in a comparison with stellar-velocity data. The definition of $\sigma$ used by \cite{2013ApJ...764..184M} involves measurements of velocity dispersion and radial velocity averaged up to some effective radius [their Eq.~(1)]. We compare their estimates with values of $\sigma$ evaluated  close to $r_H$, where the Hernquist profile is expected to give the largest contribution to their averaged estimations.}} to the measurements in the sample of large elliptical galaxies collected by \cite{2013ApJ...764..184M}.

Self-consistent (and therefore more realistic) models have also been developed to  describe photometric and kinematical data in elliptical galaxies (see e.g. \citealt{2000dyga.book.....B}) but we opted for the Hernquist profile because it reproduces the kinematical properties quite well despite   its analytical simplicity. We model the cluster DM halo with a Navarro-Frenk-White (NFW) profile  \citep{1996ApJ...462..563N,1997ApJ...490..493N}, which has been found to be in good agreement with galaxy cluster data \citep{2000AJ....119.2038V}.
The NFW mass-density profile is
\begin{align}
\rho_{\rm DM}(r)= \frac{c^3g_c \Delta_v(z)}{3} \rho_c(z)\frac{1}{(cr/r_v) \left(1+ c r/r_v\right)^2}\,,
\label{rhodm}
\end{align}
where $r_v$ is the virial radius; $\Delta_v(z)$ is the virial overdensity (see below); $c$ is a concentration parameter; the function $g_c$ is given by
\begin{equation}
g_c=\frac{1}{{\rm ln}(1+c)-c/(1+c)};
\end{equation}
and $\rho_c(z)$ is the critical density of the Universe at the redshift under consideration,
\begin{align}
\rho_c(z)=\frac{3 H^2(z)}{8\pi G},
\end{align}
where
\begin{align}
H(z)=H_0\sqrt{(1+z)^3\, \Omega_M+\Omega_\Lambda}.
\label{Hofz}
\end{align}
The virial radius $r_v$ is defined as the distance from the center of the halo within which the mean density is $\Delta_v(z)\rho_c (z)$. The halo mass $M_{\rm DM}$ is then simply defined to be the DM mass within $r_v$:
\begin{align}
&M_{\rm DM}=\frac{4}{3}\pi r_v^3  \Delta_v(z)\rho_c (z) \,.
\label{mvirdef}
\end{align} 
Under the assumption that the cluster has just virialized \footnote{For simplicity, we do not truncate the NFW halo at the virial radius, which is expected under such virialization assumption (e.g. \citealt{2003astro.ph..9240P, 2012MNRAS.423.2533B}). Our predictions of the final occupation fractions are independent of this assumptions: SMBHs kicked at $r_{\rm max}> r_v\sim {\rm few ~Mpc}$ in general do not find their way back to the galactic center within a Hubble time.}, calculations of spherical top-hat perturbations \citep{1980lssu.book.....P} yield $\Delta_v=18 \pi^2\simeq 178$, but the actual value depends on the cosmological model through \citep{1993MNRAS.262..627L,1998ApJ...495...80B,2011ApJ...740..102K}
\begin{align}
\Delta_v(z)&=18 \pi^2 - 82 \Omega_\Lambda(z)-39 \Omega_\Lambda^2(z),
\label{virialover}
\end{align}
where
\begin{align}
\Omega_M(z)&=\frac{(1+z)^3 \,\Omega_M}{(1+z)^3\, \Omega_M + \Omega_\Lambda},
\quad \Omega_\Lambda(z)=1-\Omega_M(z).
\end{align}
The virial radius as a function of the halo mass reads
\begin{equation}
{r_v}= \left(\frac{M_{\rm DM}}{10^{14} M_\odot}\right)^{1/3}\left(\frac{\Omega_M}{\Omega_M(z)}\frac{\Delta_v(z)}{18\pi^2}\right)^{-1/3}\frac{1\, \text{Mpc}}{1+z}.
\end{equation}
 In the regime considered here ($z<1 $), the virial overdensity $\Delta_v$ is roughly $0.7\times 18 \pi^2\simeq 124$ with a rather weak dependence on $z$; typical sizes of DM halos with the same mass may differ by a factor $\sim 1.5$ if placed at different redshifts.

\begin{figure}
\includegraphics[width=\columnwidth]{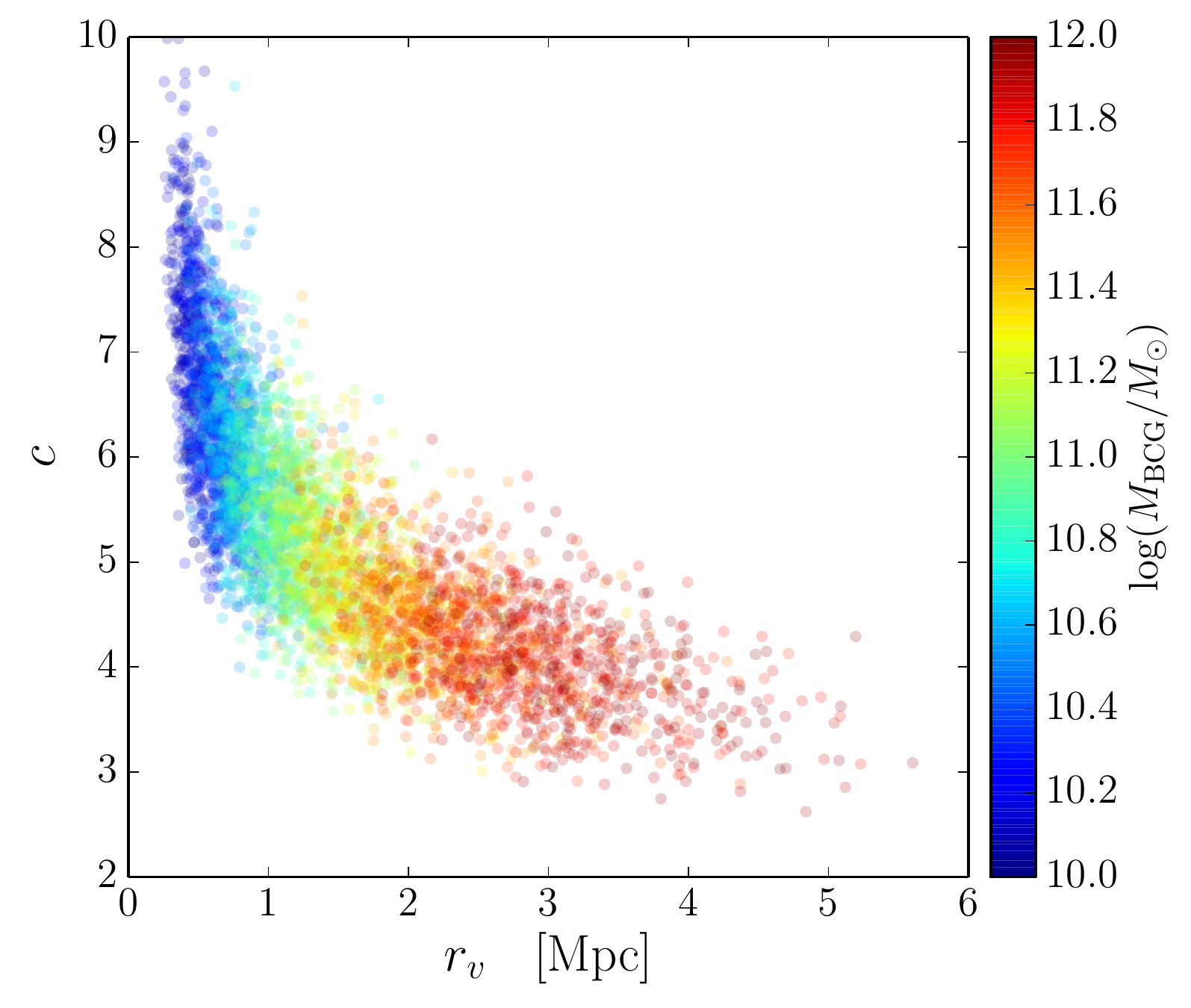}
\caption{(color online)  Observationally based relation between the halo virial radius $r_v$ and the concentration parameter $c$. Fitting formulas provided by 
\protect\cite{2012MNRAS.422.2213S} and \protect\cite{2007MNRAS.381.1450N} 
are solved using the iterative procedure described in the main text. $M_{\rm BCG}$ is reported on the color scale. Massive galaxies (lighter points on the right) correspond to larger halos and to lower values of $c$; on the other hand, lighter BCGs (darker points on the left) are hosted in smaller halos and present a wider range of concentrations up to $c\simeq 10$.  This figure is obtained with a uniform distribution in $\log{M_{\rm BCG}/M_{\odot}}\in[10,12]$ at $z=0$. }
\label{rv_c_iter}
\end{figure}

\begin{figure}
\includegraphics[width=0.95\columnwidth]{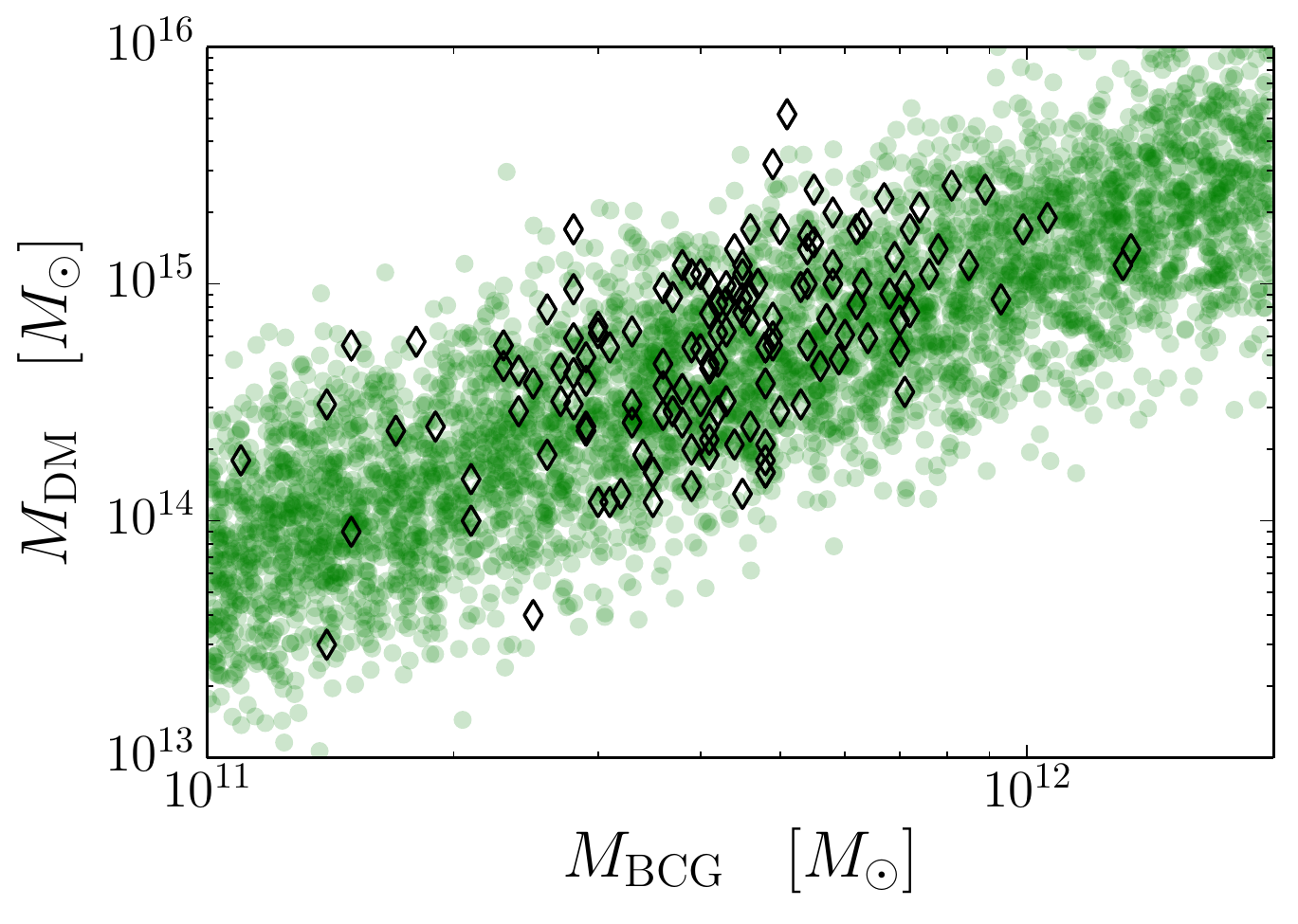}
\caption{(color online)   Relation between $M_{\rm BCG}$ and $M_{\rm DM}$ as implemented in our model. Our Monte Carlo realization (green circles) is statistically consistent with the observational catalog of 160 BCGs collected by  \protect\cite{2012MNRAS.427..550L} (black diamonds). This figure is obtained with uniform distributions in $\log{M_{\rm BCG}/M_{\odot}}\in[11,12.3]$ and $z \in [0,1.5]$, which are the same ranges covered by the data sample in  \protect\cite{2012MNRAS.427..550L}. }
\label{Lidman}
\end{figure}

\cite{2012MNRAS.422.2213S} relate the BCG visible mass to the  halo mass measured at $r_{500}$, defined to be the radius at which the mean density is 500 times the critical density of the \emph{present} Universe
\begin{align}
M_{\rm 500}=\frac{4}{3}\pi r_{500}^3 \, \rho_c(z=0) \times 500.
\label{m500def}
\end{align}
Their observational relation reads \citep{2012MNRAS.422.2213S}
\begin{equation}
{\rm log}\left(\frac{M_{\rm 500}}{10^{14}\msun}\right)=-14.29+1.28 \log\left(\frac{M_{\rm BCG}}{M_\odot}\right),
\label{m500mbcg}
\end{equation}
with dispersion $\sigma\approx0.3$~dex.  
The concentration parameter $c$ is related to the halo mass and in general depends on the redshift and the underlying cosmological model \citep{2007MNRAS.381.1450N,2008MNRAS.391.1940M,2014MNRAS.441..378L}. Those dependencies are however rather weak in the BCG range ($M_{\rm 200}\sim 10^{13-16} M_{\odot}$), in which theoretical predictions by different authors  tend to agree (see Fig.~10 in \citealt{2014MNRAS.441..378L}). 
Here we implement the  relation reported by \cite{2007MNRAS.381.1450N}
\begin{equation}
\log c=5.26-0.1\log\left(\frac{M_{\rm 200}}{10^{14}\msun}h^{-1}\right)\,,
\label{Netorel}
\end{equation}
with a dispersion of $0.05$~dex. In analogy with Eq.~(\ref{m500mbcg}), $M_{200}$ is defined to be the mass of the halo inside a radius $r_{200}$ at which the mean density is $200$ times the critical density
\begin{align}
M_{\rm 200}=\frac{4}{3}\pi r_{200}^3 \, \rho_c(z=0) \times 200.
\label{m200def}
\end{align}
The value of $M_{500}$ and $M_{200}$ can also be obtained by integrating $\rho_{\rm DM}(r)$ from Eq.~(\ref{rhodm}). This gives the following constraints on $r_{200},r_{500}$ and $r_v$:
\begin{align}
\frac{500}{\Delta_v}\frac{H_0^2}{H^2(z)}= g_c \left(\frac{r_v}{r_{500}}\right)^3 \left[ \ln \left( 1+\frac{c r_{500}}{r_v}\right) - \frac{c r_{500}/r_v}{1+ c r_{500}/r_v}\right]; 
\label{constraint500}\\
\frac{200}{\Delta_v}\frac{H_0^2}{H^2(z)}= g_c \left(\frac{r_v}{r_{200}}\right)^3 \left[ \ln \left( 1+\frac{c r_{200}}{r_v}\right) - \frac{c r_{200}/r_v}{1+ c r_{200}/r_v}\right].
\label{constraint200}
\end{align}
We implement an iterative procedure to find $r_v$ and $c$ simultaneously; results are presented in Fig.~\ref{rv_c_iter}.
For each BCG stellar mass, $M_{\rm BCG}$, we compute $M_{500}$ trough Eq.~(\ref{m500mbcg})  assuming a Gaussian error of 0.3~dex, and then $r_{500}$ using Eq.~(\ref{m500def}). Given the initial guess $c=5$, the constraint (\ref{constraint500}) is used to obtain numerically $r_v$. Eq.~(\ref{constraint200}) is then solved to find $r_{200}$, and $M_{200}$ is obtained using Eq.~(\ref{m200def}). An updated value of $c$ can now be computed through the observational relation (\ref{Netorel}). The whole procedure is then iterated. When convergence is reached\footnote{Convergence down to $|\Delta c|<10^{-6}$ is typically obtained after 5 iterations.}, we add a Gaussian error of 0.05~dex to the final value of $c$. 
Once $r_v$ and $c$ are obtained, the halo mass, $M_{\rm DM}$, is given by Eq.~(\ref{mvirdef}). As a consistency test, the BCG/DM-halo relation is shown in Fig.~\ref{Lidman}, where our Monte Carlo sample is contrasted to observational data from \cite{2012MNRAS.427..550L}.  

To summarize: we model the BCG mass density from Eqs.~(\ref{rhobcg}) and (\ref{rhodm}) as $\rho=\rho_{\rm BCG}+\rho_{\rm DM}$, while the associated gravitational potential is given by $\Phi=\Phi_{\rm BCG}+\Phi_{\rm DM}$, with
\begin{align}
\Phi_{\rm BCG}(r)&=-\frac{GM_{\rm BCG}}{r+r_H},
\end{align}
and
\begin{align}
\Phi_{\rm DM}(r)&=-g_c\frac{GM_{\rm DM}}{r_v}\frac{{\rm ln}(1+c {r}/{r_v})}{r/r_v}.
\label{phidm}
\end{align}

%%%%%%%%%%%%%%%%%%%%%%%%%%%%%%%%%%%%%%%%%%
\subsection{Recoiled SMBH return timescales} 
\label{timebackmodel}

Following the binary merger, the remnant SMBH recoils because of asymmetrical GW emission which may result in its ejection from the BCG core. The recoiling SMBH transfers its orbital energy into random motions of the surrounding stars through collisions,  and may sink back to the galactic center.  Here we develop two physical models to predict the return timescale of this process.

The remnant SMBH is initially kicked out on a radial orbit. Detailed N-body simulations of the process have been performed by \cite{2008ApJ...678..780G}, which detect strong damping during each passage of the SMBH though the galactic core. It is therefore critical to know whether the recoiling SMBH orbit crosses the galactic core, since damping happens mainly in those quick passages. 
Repeated core passages cannot be prevented in a spherically symmetric potential. However, post-merger galactic potentials are  expected to be  triaxial \citep{2011ApJ...732L..26P,2011ApJ...732...89K}: the SMBH orbit will not in general remain exactly radial and in particular the core may not be crossed \citep{2007ApJ...662..797V}. Moreover, especially for extreme kicks, the SMBH can travel further than a Mpc from the BCG core. At this point its trajectory is likely to be perturbed by the clumpy potential of other galaxies and DM subhalos within the main cluster halo, and return to the BCG core is unlikely. Missing the core would result in a much longer inspiral timescale because only low-density regions contribute to the frictional force.
This difference is critical to our purposes, particularly if this timescale gets comparable with the timescale between two galactic mergers: less efficient sinking may result in "empty" galactic centers when the next satellite galaxy merges into the BCG.
The full complexity of the problem cannot be solved within our spherically symmetric model; therefore, we developed two extreme approaches bracketing the uncertainties related to the dynamics describe above. 
\begin{enumerate}
\item 
In the first model, we assume that the SMBH orbit is "quasi-circular" and we compute the sinking timescale using \citeauthor{1943ApJ....97..255C}'s (\citeyear{1943ApJ....97..255C}) dynamical friction (DF). This is meant to be the extreme case for a strongly perturbed potential for which the SMBH  never crosses the galactic core.
\item
 In the second scenario, we consider repeated SMBH-core bounces by fitting the N-body  simulations reported by \cite{2008ApJ...678..780G}. This model is appropriate for BCG and cluster potentials which exhibit small deviations from spherical symmetry.
 \end{enumerate}
 \begin{figure*}
\includegraphics[width=0.495\textwidth]{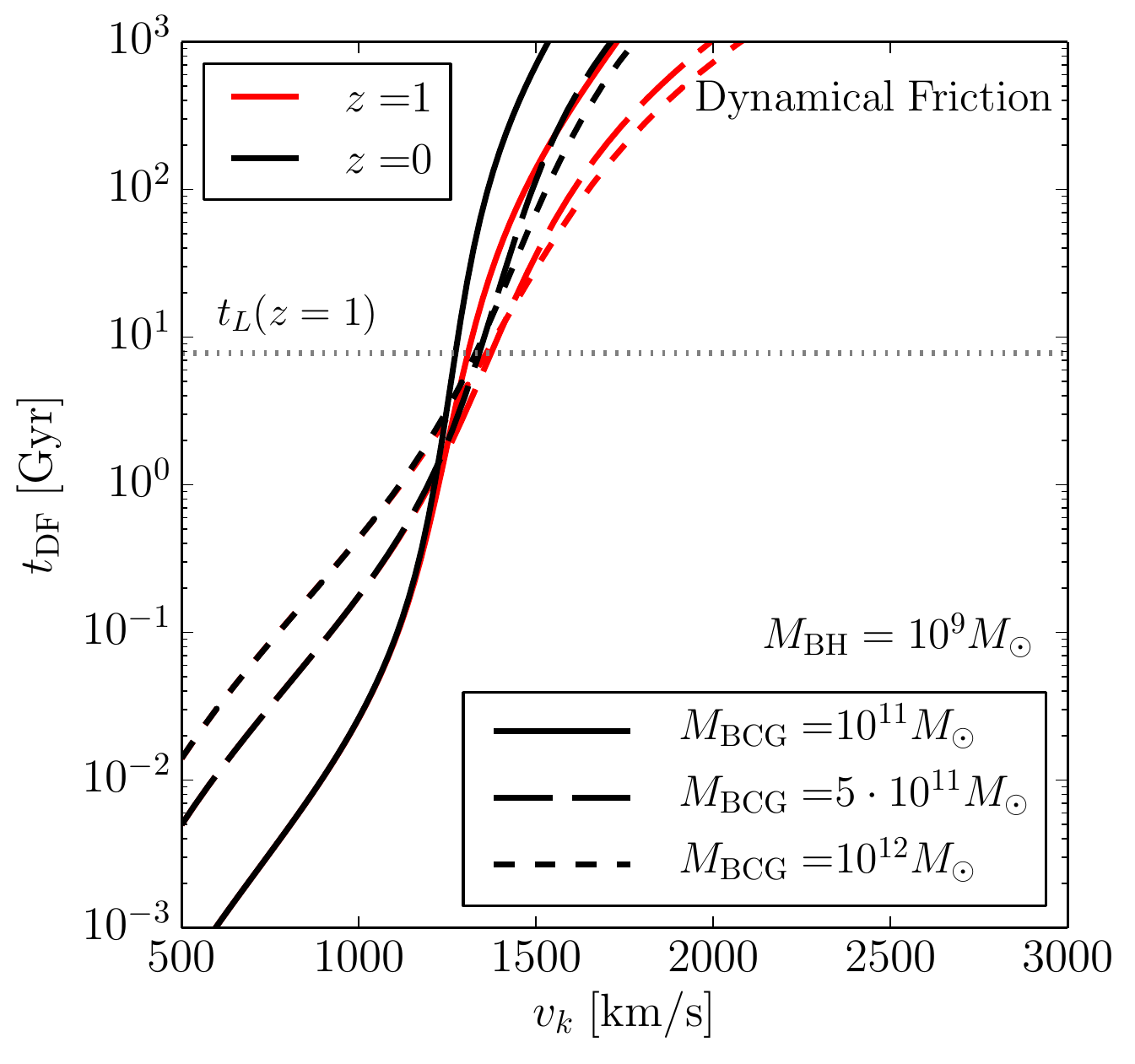}
\includegraphics[width=0.495\textwidth]{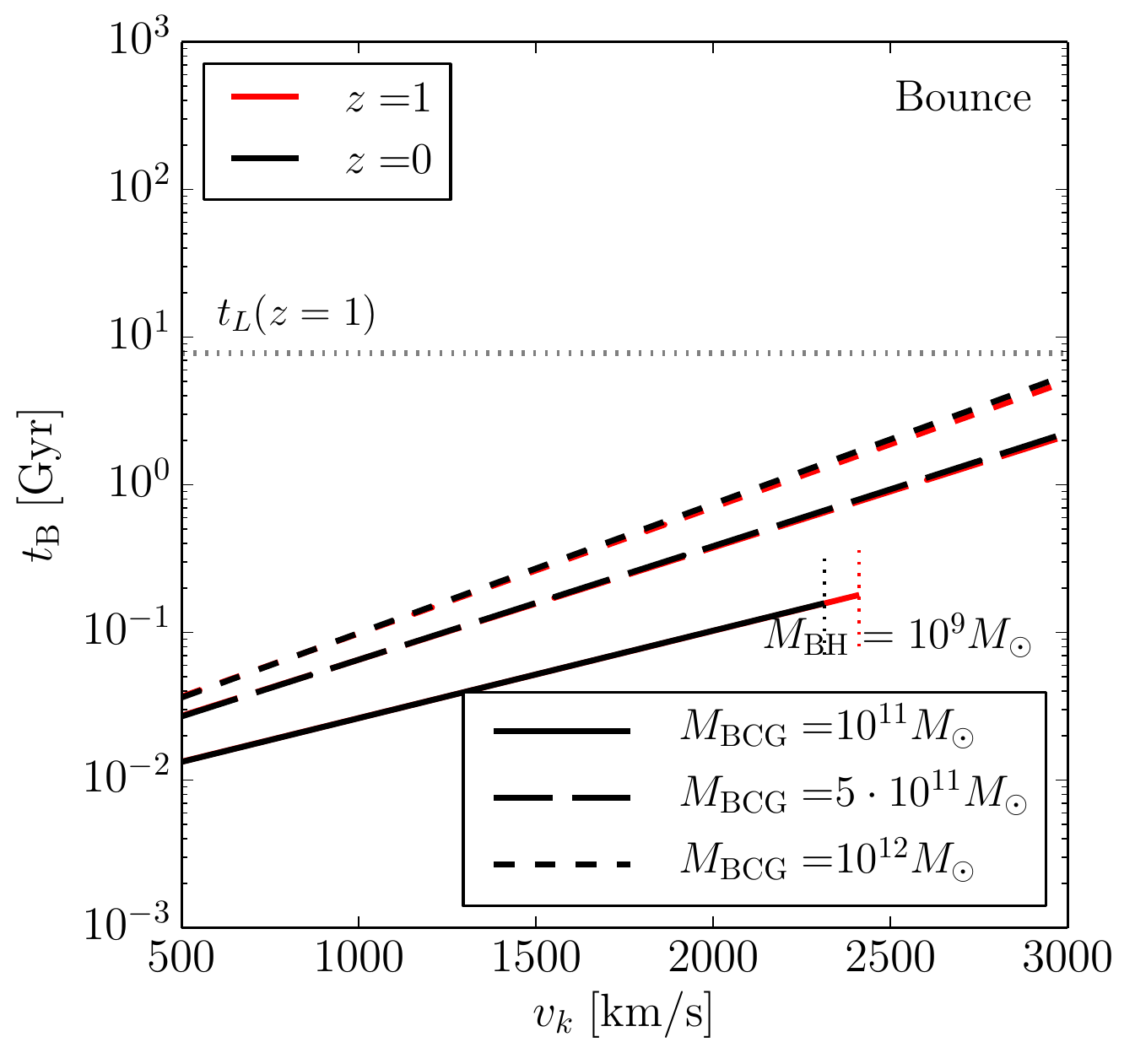}
\caption{(color online) SMBH return timescales, in both the DF (left) and the bounce model (right), as a function of the kick velocity $v_k$. We consider recoiling SMBHs with $M_{\rm BH}=10^{9} M_{\odot}$ and BCGs with stellar mass $M_{\rm BCG}=10^{11} M_{\odot}$ (solid), $5\cdot10^{11} M_{\odot}$ (long-dashed) and $10^{12} M_{\odot}$ (short-dashed). The remaining galaxy parameters (such as $r_H$, $M_{\rm DM}$, $r_v$ and $c$) are estimated using the prescriptions presented in Sec.~\ref{bcgmodel}. To facilitate comparisons,  here we set variances in  Eqs. (\ref{breakradius}), (\ref{m500mbcg}) and (\ref{Netorel}) to zero. In order to bracket the effects of cosmological evolution we carry out the analysis at both $z=0$ (darker, black lines) and $z=1$ (lighter, red lines).
BHs are effectively ejected from the BCGs when the sinking timescale (either $t_{\rm DF}$ or $t_{\rm B}$) gets larger than the lookback time at the merger redshift, which in turn is always smaller than the one computed  at $z=1$ ($\sim 7.8$ Gyr, shown with a dotted horizontal line). Dotted vertical lines in the right panel are placed at the escape velocity $v_{\rm esc}$, at which Eq.~(\ref{myfit}) must be truncated.}
\label{kick_df}
\end{figure*}

%%%%%%%%%%%%%%%%%%%%%%%%%%%%%%%%%%%%%%%%%%
\subsubsection{Dynamical-friction model}

 Let us consider a SMBH with mass $M_{\rm BH}$ kicked with velocity $v_k$ from the galactic center ($r=0$). The SMBH will be ejected from the galactic halo if $v_k$ exceeds the escape velocity of the system
\begin{align}
v_{\rm esc}= \sqrt{2G\left( \frac{M_{\rm BCG}}{r_H} +  c \,g_c \frac{M_{\rm DM}}{r_v}\right)} \,.
\label{escapespeed}
\end{align}
If $v_k<v_{\rm esc}$, the SMBH will stop at a distance $r_{\rm max}$ from the center.
\cite{2008ApJ...678..780G} showed that  the maximum displacement $r_{\rm max}$ can be estimated simply trough energy conservation neglecting star friction (see their Fig.~2)
\begin{align}
\frac{1}{2} v_k^{2}+\phi(0)=\phi(r_{\rm max})\,.
\label{stoppoint}
\end{align}
The initial displacement is reached in a time which is typically 100 times smaller than the sinking timescale \citep{2008ApJ...678..780G} and will be therefore neglected.
Here we estimate the time needed to sink back to $r=0$ integrating the DF equation on quasi-circular orbits. 
The frictional force exerted onto the BH is given by (e.g. \citealt{1987gady.book.....B})
\begin{align}
F(r)=\frac{4 \pi G^2 M_{\rm BH}^2 \rho(r)\,\xi(r) \ln \Lambda}{v_c^2(r)}\,,
\label{frictionalforce}
\end{align}
where $v_c(r)=\sqrt{r\, d\phi /dr}$ is the circular velocity, $\ln\Lambda$ is the Coulomb logarithm and the factor $\xi(r)$ depends on the stellar velocity distribution.
We take $\ln \Lambda=2.5$, as observed by \cite{2008ApJ...678..780G} in the very first phase of their simulated orbits (see also \citealt{2004ApJ...607..765E}). 
We assume the velocity distribution to be locally Maxwellian, with velocity dispersion $\sigma(r)$. Although not exact, the Maxwellian  distribution is approached as a consequence of collisionless relaxation processes \citep{1967MNRAS.136..101L}. 
Under this assumption, the $\xi$ factor in Eq.~(\ref{frictionalforce}) reads \citep{1987gady.book.....B}
\begin{align}
\xi(r)&= \text{erf}\left[\frac{v_c(r)}{\sqrt 2 \sigma(r)}\right] - \sqrt{\frac{2}{\pi}}\frac{v_c(r)}{\sigma(r)}\exp\left[-\frac{v_c^2(r)}{2\sigma^2(r)}\right]\,.
\end{align}
The velocity dispersion $\sigma(r)$ is computed from our galactic potential using the expression provided by \cite{1980MNRAS.190..873B} when isotropy is assumed.
The frictional force $F(r)$ is tangential and directed opposite to the SMBH velocity. The SMBH angular momentum \mbox{$L(r)=M_{\rm BH} r v_c(r)$} is lost at the rate \mbox{$dL(r)/dt=-r F(r)$} by Newton's third law, causing the SMBH to slowly inspiral while remaining on a quasi-circular orbit. The DF timescale, over which the SMBH sinks back to the galactic center $r=0$ from its initial position $r_{\max}$, is thus given by\footnote{Because of the intrinsic divergence in the density profile (\ref{rhobcg}-\ref{rhodm}), this integral cannot be computed up to $r=0$: hereafter, we implement a lower threshold at $10^{-3} r_H \sim 1 {\rm \, pc}$.  We also neglect the dependence on the redshift while computing the integral (\ref{DFtimescale}). In both models, the sinking times are computed fixing the redshift at his initial value (i.e. when the kick is imparted to the SMBH). As shown in Fig.~\ref{kick_df},  differences between timescales computed at different redshifts are  negligible in the interesting region $t_{\rm DF}<t_L(z=1)$.}
\begin{align}
t_{\rm DF}=-\int_{r_{\rm max}}^0 \frac{d L(r)}{dr} \frac{1}{r F(r)} dr\,.
\label{DFtimescale}
\end{align}

DF timescales for typical systems are reported in Fig.~\ref{kick_df} (left panel) as a function of the kick velocity $v_k$.
A recoiling SMBH is strictly ejected only if $v_k>v_{\rm esc}$, which is unlikely since we are considering the whole cluster potential for which $v_{\rm esc}$ may be as large as $\sim 6000$ Km/s for the typical values $M_{\rm BCG}=10^{12} M_{\odot}$ and $M_{\rm BH}=10^9 M_{\odot}$. However, SMBHs are effectively ejected if their return timescales are larger than the lookback time at the merger redshift $z_m$ (e.g. \citealt{1993ppc..book.....P})
\begin{align}
t_L(z_m)=\int_0^{z_m} \frac{dz}{(1+z) H(z)}\,, 
\label{lookback}
\end{align}
which corresponds to the  time the Universe needs to evolve from $z_m$ to now. In this case, the SMBH remains outside the BCG, wandering in the intracluster medium. Our systems are evolved from $z=1$ to $z=0$, which sets a (conservative) effective escape condition $t_{\rm DF}>t_L(z=1)$ for which SMBHs will never come back to the BCG center. As shown in the left panel of Fig.~\ref{kick_df}, this condition is fulfilled for achievable kicks $v_k\sim 1500$ Km/s, opening the possibility of several (effective) ejections from typical BCGs.
When this occurs, the distance between the SMBH and the galaxy center (offset) can be estimated by numerically inverting Eq.~(\ref{DFtimescale}). At $z=0$, the SMBH  needs the additional time $t_{\rm DF}-t_L(z_m)$ to sink to the center. The offset $r_{\rm z=0}$ is given by the displacement resulting in such time\footnote{In both scenarios, offsets are computing with the galaxy properties at $z=0$.}, i.e.
\begin{align}
t_{\rm DF}-t_L(z_m)=-\int_{r_{\rm z=0}}^0 \frac{d L(r)}{dr} \frac{1}{r F(r)} dr\,.
\label{}
\end{align}

%%%%%%%%%%%%%%%%%%%%%%%%%%%%%%%%%%%%%%%%%%

\subsubsection{Bounce model}

To describe recoiling SMBHs on radial orbit, we rely on the N-body simulations performed by  \cite{2008ApJ...678..780G}. They study the motion of a SMBH recoiling from the center of an initially spherically symmetric galaxy. The SMBH motion can be divided into three distinct stages: (i) firstly, a short DF phase damps the radial oscillations as predicted by \citeauthor{1943ApJ....97..255C}'s (\citeyear{1943ApJ....97..255C}) formula with $2\lesssim\ln\Lambda \lesssim 3$; (ii) once the amplitude of the motion is smaller than the core radius, the SMBH and the galactic core exhibit oscillations about their common center of mass; (iii) finally, the SMBH and the core reach thermal equilibrium when the SMBH kinetic energy equals the mean kinetic energy of the stars in the core. Orbital energy dissipation occurs mostly during core-SMBH encounters. Here we are interested in estimating the timescale $t_{\rm B}$, given by the sum of the first- and the second-phase.

The duration of the first two phases is listed in \cite{2008ApJ...678..780G} for 18 simulations in total, 6 in each of their 3 different models. As suggested by the authors themselves [their Eq.~(18)],  the second-phase times originally reported must be corrected, since the number of N-body particles used is smaller than the actual number of stars in a galaxy.  They implement the galaxy profile firstly proposed by \cite{2005MNRAS.362..197T} to describe binary-depleted galactic cores which present a well defined profile transition at the core radius $r_c$. Oscillations damp only during passages through the galaxy core, whose properties are expected to strongly influence the damping time. For a given $M_{\rm BCG}$, we firstly compute the SMBH mass $M_{\rm BH}$, the velocity dispersion $\sigma_c$ and the mass density $\rho_c$ at $r_c$ for each of their three models using the \cite{2005MNRAS.362..197T} density profile. 
Even if DF cannot fully describe such core-passage dynamics,  the return time  appear to satisfy the same scaling relation as if DF would be fully responsible for the sinking process \citep{2008ApJ...678..780G}.
We therefore scale the simulated kick velocities with $\sigma_c$ and the reported return timescales $t_{\rm B}$  with  ${\sigma_c^3}/{G^2 \rho_c M_{\rm BH}}$. 

\begin{figure}
\includegraphics[width=\columnwidth]{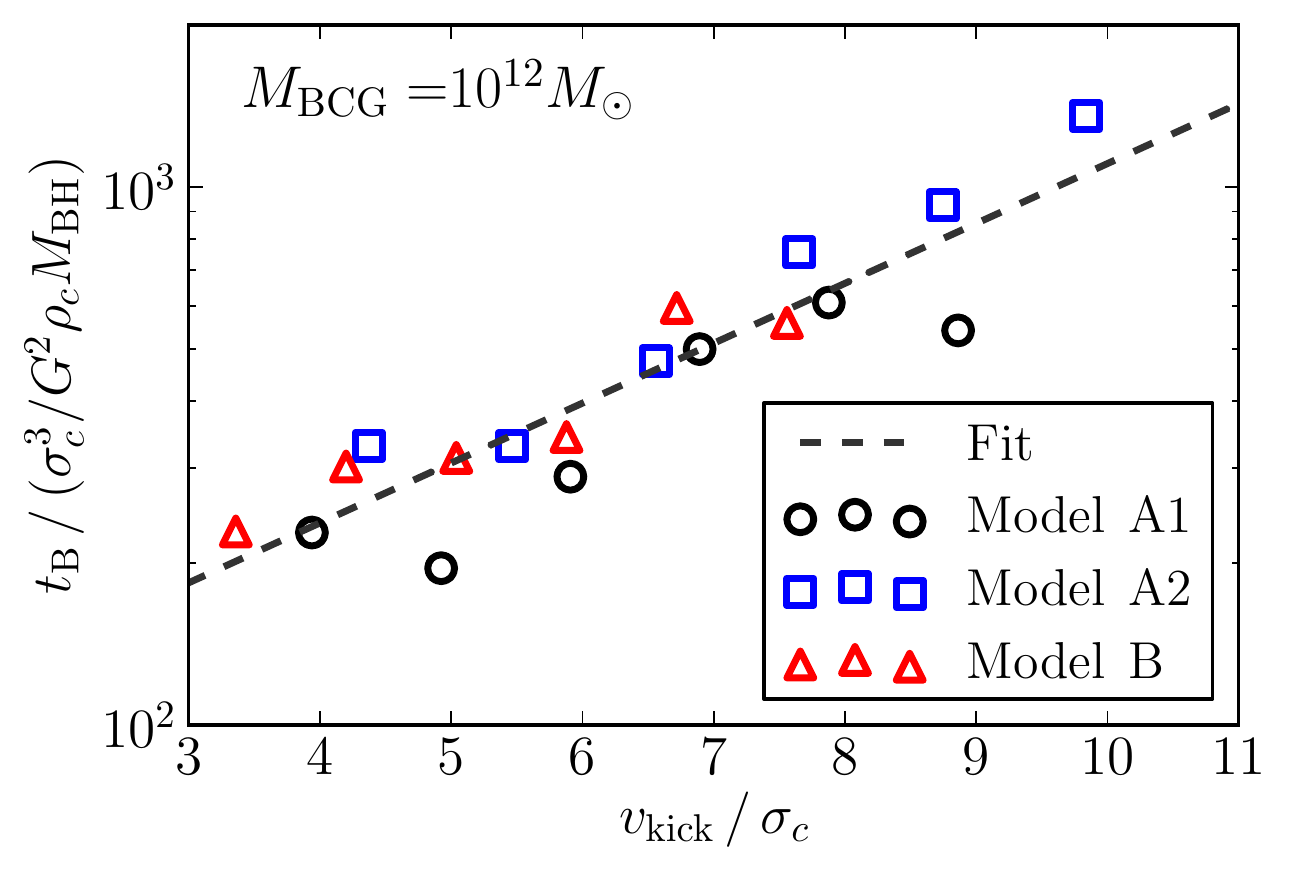}
\caption{(color online) Fitting curve employed to compute the return timescale in the bounce model  $t_{\rm B}$ as a function of the kick velocity ${v_{\rm kick}}$. Markers show predictions computed by \protect\cite{2008ApJ...678..780G} in each of their models, namely \textit{A1}, \textit{A2} and \textit{B}. Once reduced to dimensionless quantities with the expected scaling, all  three models appear to lie on the same lin-log relation, which however must be truncated at the escape velocity $v_{\rm esc}$. The dimensionless-scaled points and the fitting curve (dashed black line) depend only weakly  on the galaxy mass $M_{\rm BCG}$. This figure is produced with $M=10^{12}\msun$; the resulting fitting coefficients are $ a=0.26$ and $b=4.44$.}
\label{GM08_fit}
\end{figure}

Once reduced to a dimensionless problem, we fit their 18 simulated timescales with the ansatz 
\begin{align}
t_{\rm B}= 
\frac{\sigma_c^3}{G^2 \rho_c M_{\rm BH}}\; \exp\left( a \frac{v}{\sigma_c} +b \right) \,,
\label{myfit}
\end{align}
truncated at the escape velocity $v_{\rm esc}$.
Here $a$ and $b$ are best fit coefficients. They only depend (weakly) on the galactic mass $M_{\rm BCG}$ which  enters in the correction factor to $t_{\rm B}$ due to the limited number of N-body particles. Fig.~\ref{GM08_fit} shows the results of our fit for a fiducial mass $M_{\rm BCG}=10^{12} \msun$.
The dimensionless fit can be reported into physical units by computing $\sigma_c$ and $\rho_c$ for our galactic profiles (Hernquist+NFW) at a fiducial core radius
\begin{align}
\log\left(\frac{r_c}{\text{pc}}\right) \simeq 1.1+0.09 \log\left(\frac{r_H}{\text{pc}}\right),
\end{align}
as obtained by matching the mass dependencies in Eq.~(\ref{breakradius}) with the analogous estimate  for the core radius used by \cite{2008ApJ...678..780G}. 
Results of our procedure are reported in the right panel of Fig.~\ref{kick_df}. This second model predicts longer inspiral timescales for kicks smaller than $\sim 1000$ Km/s; while large kicks make SMBHs returning very quickly ($\sim 100$ Myr) to their galactic centers. 
If the SMBH does not escape from the cluster ($v<v_{\rm esc}$), there will always be a first core passage causing enough dissipation to trigger more and more passages leading to a quick comeback.

The SMBH offset at $z=0$ can be computed by iterating the fit procedure describe above. We numerically look for the hypothetical kick velocity $\tilde v_k$ which would result in a return time equal to $t_B - t_L(z_m)$, i.e. the time left to the SMBH at $z=0$ to finally reach the galactic center.  Assuming the SMBH motion to be approximately oscillatory, we compute the amplitude of the oscillations $\tilde r_{\rm z=0}$ from energy conservation [cf. Eq.~(\ref{stoppoint})] and we finally estimate the offset to be $r_{z=0} =  \tilde r_{\rm z=0} \sin \varphi$, with $\varphi$ uniformly distributed in $[0,\pi]$.

%%%%%%%%%%%%%%%%%%%%%%%%%%%%%%%%%%%%%%%%%%
\subsection{BCG merger rates}
\label{mergermodel}

In the last few years, strong observational evidence for a prominent growth of BCGs from $z=1$ came about. Among other studies, \cite{2011MNRAS.415.3903T} observe that early-type galaxies grew by a factor 5-10 in size and 2-4 in mass since $z=1$, and \cite{2012MNRAS.427..550L} find that BCGs grow in mass by a factor of $\approx 2$ in the redshift range $0.9-0.2$ (see also \citealt{2013MNRAS.434.2856B} and \citealt{2014MNRAS.442..589A}). BCG mass growth is naturally explained by frequent mergers in the hierarchical build-up scenario, and several dedicated simulations and theoretical studies find that major and minor mergers can account for it \citep{2007MNRAS.375....2D,2010ApJ...725.2312O,2012MNRAS.425..641L,2013MNRAS.435..901L}. However, there are claims that size growth cannot be ascribed to mergers, and might be related to the redshift evolution of the properties of the underlying dark matter halos \citep{2014ApJ...786...89S,2014MNRAS.440..610P}. In general, the merger-driven mass-growth scenario is consistent with observations of close galaxy pairs \citep{2009MNRAS.396.2003L,2009ApJ...697.1369B,2009A&A...498..379D,2010ApJ...719..844R,2012ApJ...747...85X,2012A&A...548A...7L}, and both observations and simulations point toward high merger rates for early-type galaxies \citep{2010ApJ...724..915H,2011ApJ...742..103L}, that can be up to 0.4/Gyr at $z\sim 1$ for BCGs \citep{2013MNRAS.433..825L}.

Here we exploit the observationally based approach put forward by \cite{2013MNRAS.433L...1S}. We are not interested in a global galaxy-merger rate, but rather in the distribution of mergers experienced by the typical BCG. Building on the same formalism as in \cite{2013MNRAS.433L...1S}, the galaxy merger rate per unit mass ratio\footnote{We indicate galaxy mass ratios with $Q$, to differentiate with black-holes mass ratios $q$.} and redshift experienced by a galaxy of a given mass can be written as:
\begin{equation}
\frac{d^2N}{dzdQ}\bigg\rvert_M=\frac{df}{dQ}\bigg\rvert_{M,z}\frac{1}{\tau(z,M,Q)}\frac{dt_L}{dz}.
\label{galmrate}
\end{equation}
Here, $df/dQ|_{M,z}$ is the differential fraction of galaxies with mass $M$ at redshift $z$ paired with a secondary galaxy having a mass ratio in the range $[Q, Q+\delta{Q}]$; $\tau(z,M,Q)$ is the typical merger timescale for a galaxy pair with a given $M$ and $Q$ at a given $z$; and $dt_l/dz$ is the integrand in Eq.~(\ref{lookback}). $df/dQ$ can be directly measured from observations, whereas $\tau$ can be inferred by detailed numerical simulations of galaxy mergers. The number of mergers experienced from $z=1$ to $z=0$ by a galaxy starting with mass $M_{\rm BCG}=M_{\rm z=1}$ at $z=1$ can be therefore written as
\begin{equation}
N({M_{\rm z=1}})=\int_1^0dz\int_{Q_{\rm min}}^1 dQ\int dM \frac{d^2N}{dzdQ}\bigg\rvert_M\delta[M-M(z)],
\label{Nmerger}
\end{equation}
where the integral is consistently evaluated at the redshift-evolving galaxy mass $M(z)$ through the Dirac delta function.

{To estimate the mass growth of BCGs, we consider the fraction $f$ of galaxies with a companion in the range \mbox{$Q_{\rm min}=0.25<Q<1$}, which correspond to the standard definition of major mergers. $f$ is estimated in several observational studies, and it is generally fitted with a function of redshift of the form} 
\begin{equation}
f=a(1+z)^b.
\label{fpair}
\end{equation}
{The parameters $a$ and $b$ are, in general, function of the primary galaxy mass. Since we are concerned with BCGs, we consider fits to Eq.~(\ref{fpair}) corresponding to primaries with mass $M>10^{11}\msun$. We construct three models, to which we will refer as "Optimistic", "Fiducial" and "Pessimistic". In the "Fiducial" model we take the best fit to the observations of \cite{2009ApJ...697.1369B}, yielding $a=0.035$, $b=1.3$. Those data are consistent with a larger fraction described by $a=0.07$, $b=0.7$, which we take as "Optimistic" model. \cite{2012A&A...548A...7L} find a smaller pair fraction with a stronger redshift dependence, corresponding to $a=0.02$, $b=1.8$, which we take as "Pessimistic" model. Pairs are then distributed across the allowed mass ratio range according to $df/dQ|_{M,z}\propto Q^{-1}$ \citep{2011A&A...530A..20L}. \cite{2012A&A...548A...7L} additionally provide the pair fraction in the range $0.1<Q<0.25$, corresponding to minor mergers. This is found to be $f\approx0.06$ independent on redshift. We add those to the "Pessimistic" model to construct the "Pessimistic-Minor" model, which we use to assess the impact of minor mergers on our findings.}

{The function $\tau$ is then specified by using the formula given by \cite{2008MNRAS.391.1489K}  [their Eq.~(10)]  to get\footnote{We fixed $r_p=30$ kpc in Eq.~(10) of \cite{2008MNRAS.391.1489K}, because this is the projected separation of the samples we use.}}
\begin{equation}
\tau = 1.32\,{\rm Gyr}\left( \frac{M_*}{4\times10^{10} h^{-1} M_\odot} \right)^{-0.3}\left(1+\frac{z}{8}\right),
\label{tau}
\end{equation}
{where $M_*$ is the total mass of the pair. We shall stress here that Eq.~(\ref{tau}) provides the {\it galaxy} merger timescale, which can be regarded as the timescale over which a bound SMBH binary forms. The actual coalescence of the binary might be further delayed because the system needs to get rid of its energy and angular momentum in order to get to the efficient GW emission stage. This is known as the "final parsec problem" \citep{2003AIPC..686..201M}; we will return on this potential caveat in the next section. The galaxy merger rate is finally obtained by inserting Eq.~(\ref{fpair}) --distributing the pairs according to $Q^{-1}$-- and Eq.~(\ref{tau}) into Eq.~(\ref{galmrate}).}

\begin{figure}
\includegraphics[width=\columnwidth]{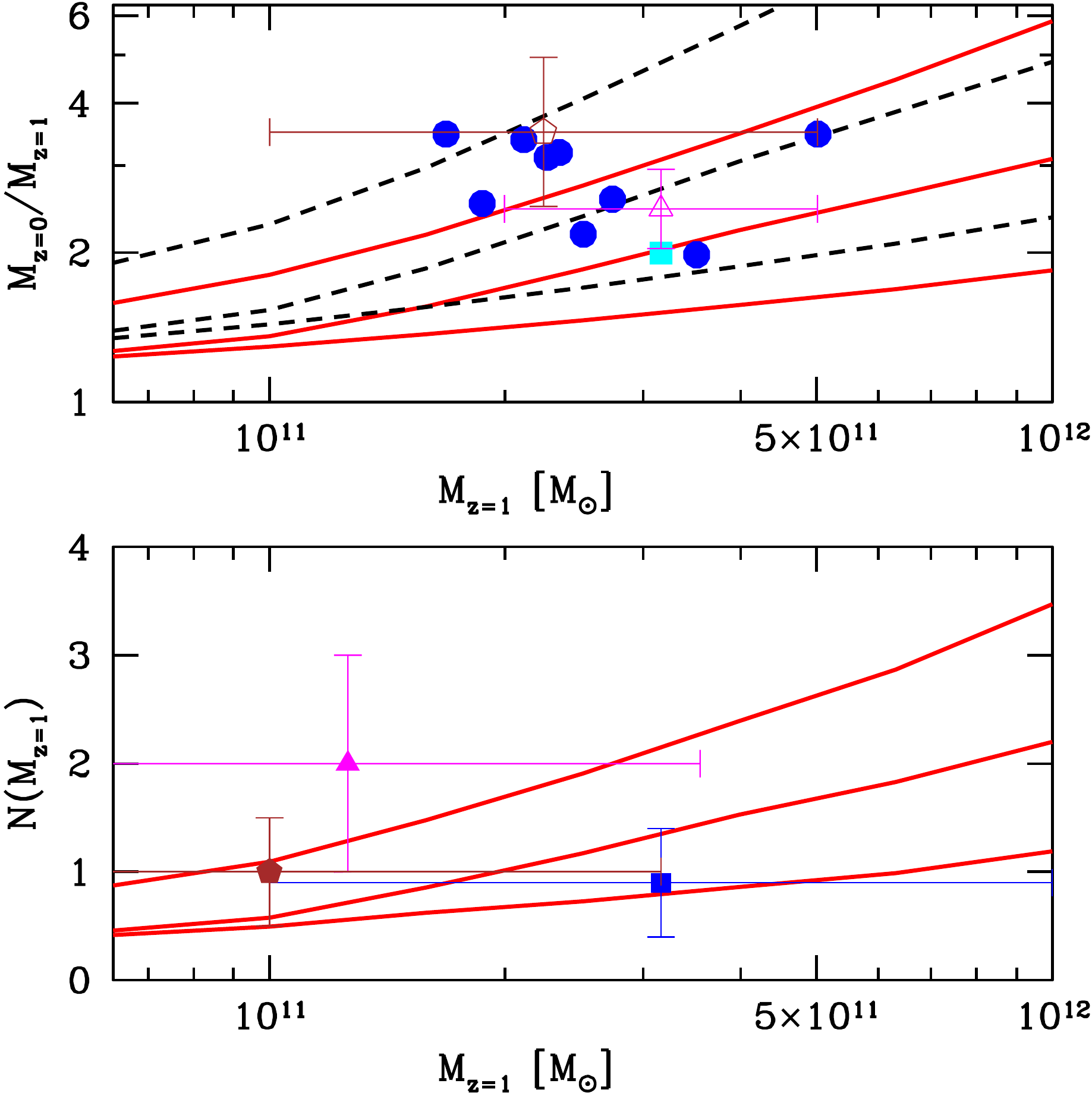}
\caption{{(color online) BCG mass growth (top panel) and average number of major mergers (bottom panel) as a function of initial mass at \mbox{$z=1$}. In both panels, red solid curves are predictions of our observation-based semianalytic models; from bottom to top: "Pessimistic", "Fiducial" and "Optimistic". In the top panel, the additional black--dashed lines are (in the same order) growth factors corrected for the contribution of minor mergers (the lower one correspond to the "Pessimistic--minor" model, whereas the same fractional growth correction factor is applied to get the other two curves). The magenta triangle is the average mass growth predicted by \protect\cite{2012MNRAS.427..550L}, the brown pentagon is derived from \protect\cite{2007MNRAS.375....2D}, the blue circles are a selected sample of BCGs from \protect\cite{2013MNRAS.435..901L}, and the cyan square is a simulation from \protect\cite{2010ApJ...725.2312O}. In the bottom panel, only the number of major mergers is considered, and we additionally plot the average number of mergers found by \protect\cite{2006ApJ...652..270B} (magenta triangle), \protect\cite{2012ApJ...747...85X} (brown pentagon) and \protect\cite{2010ApJ...724..915H} (blue square).}}
\label{massgrowth}
\end{figure}

{Fig.~\ref{massgrowth} compares the predicted mass growth and average number of mergers suffered by BCGs as a function of their mass at $z=1$ to a number of observations and theoretical models. When corrected for the expected contribution of minor mergers, the ``Fiducial'' model predicts a mass growth in line with observations by \cite{2012MNRAS.427..550L}. The ``Optimistic'' one has a larger growth, consistent with theoretical modelling by \cite{2007MNRAS.375....2D} and \cite{2013MNRAS.435..901L}, whereas the ``Pessimistic'' is marginally consistent with the data, and tends to slightly underpredict the BCG mass growth (still yielding to mass doubling since $z=1$). We will consider all models in the following, and we stress that our main results do not qualitatively depend on the details of the growth history of BCGs, so long as most galaxies experience at least one merger at $z<1$. }

A small fraction of our galaxies can grow up to $10^{13}\msun$ (in the ``Optimistic'' scenario in particular), which might be at odd with the sharp cutoff in the galaxy mass function observed around $10^{12}\msun$ \citep{2003ApJS..149..289B}. However, determinations of the mass function are typically obtained by converting luminosities to stellar masses. This results in large systematic uncertainties (especially at the high-mass end) due to the assumptions on the stellar mass-to-light ratio, as well as the different possible light profile fitting procedures \citep{2010MNRAS.404.2087B}, which can extend the high mass tail of the galaxy mass function by 0.5~dex \citep{2013MNRAS.436..697B}. Moreover, extreme cases of BCGs with masses possibly in excess of $5\times10^{12}\msun$ have been reported, the most notable case being ESO 146-IG 005 \citep{2010ApJ...715L.160C}.

%%%%%%%%%%%%%%%%%%%%%%%%%%%%%%%%%%%%%%%%%%
\subsection{Putting the pieces together}
\label{sec2.5}

We select the initial BCG mass at $z=1$ using the high-redshift sample collected by \cite{2012MNRAS.427..550L}, consisting in 32 observed BCGs with redshift within 0.8 and 1.6. For each initial galaxy of mass $M_{\rm BCG}=M_{\rm z=1}$, we assign a number of mergers drawn from a Poissonian distribution with average  $N({M_{\rm z=1}})$; mass ratios  and redshifts of galactic mergers are distributed according to $dN/dzdQ$ as reported in Eq.~(\ref{galmrate})\footnote{We bin mass and merger distributions and we generate our Monte Carlo samples accordingly. Bin widths have been determined through numerical experiments: 10 bins have been used to map the BCG mass distribution from the \cite{2012MNRAS.427..550L} data; 5 bins have been considered to obtain the average merger numbers $N({M_{\rm z=1}})$ (a Poissonian dispersion is then applied), while for $dN/dzdQ$ we used 4 bins in the mass ratio and 37 bins in the redshift (bin widths are smaller for $z<0.3$,  where redshifts get closer to the end of the simulations $z=0$).}. Both BCG and each satellite galaxy, are then populated with SMBHs using the SMBH-bulge relation  as recently obtained by \cite{2013ApJ...764..184M}
\begin{align}
\log \left(\frac{M_{BH}}{M_\odot}\right)=8.46+1.05\log \left(\frac{M_{BCG}}{10^{11}M_\odot}\right) \,, 
\label{mbhmgal} 
\end{align}
with a dispersion of $0.34$ dex. In particular, \cite{2013ApJ...764..184M} detect steeper slopes in the galaxy scale laws when BCG data are included in the fitted sample \citep[cf. also][]{2013ARA&A..51..511K}. When a BGC merges with a satellite galaxy, we assume that the satellite mass is fully accreted by the BCG 
\begin{align}
M'_{\rm BGC}=(1+Q)M_{\rm BCG}.
\end{align}
and we compute the stellar and DM profile from $M'_{\rm BGC}$ using the procedure described in Sec.~\ref{bcgmodel}.
No SMBH remnant can be present in the post-merger BCG  if both the parent BCG and satellite did not host any SMBH at their centers; a single SMBH is assumed to lie in the newly formed BGC if only one of the parents carried a SMBH; finally, if both the BCG and the satellites had a SMBH, we assume that the two SMBHs also merge at the same time (redshift) as the galaxies merge. 
At each SMBH merger, we compute the remnant mass, spin and recoil as presented in Sec.~\ref{bhmergermod}. From the kick velocity and the galactic potential of the newly formed BCG, we compute the return time $t_{R}$ using either $t_{\rm DF}$ from Eq.~(\ref{DFtimescale}) or $t_B$ from Eq.~(\ref{myfit}) in each of our two models. In practice, the SMBH is removed from the simulation and placed back to the galactic center after a time $t_R$. If $t_R$ is smaller than the time between two galactic mergers, the SMBH will simply settle back at the center of its BCG; if instead a subsequent galactic merger happens before, the BCG center may already contain a SMBH (coming from one of the satellites). A new binary merger is computed, possibly resulting in another ejection from the BCG.

\subsection{Possible caveats}
A few simplifying assumptions have been made in the implementation of this procedure, which we justify in the following.
 
Firstly, we assume that all SMBH binaries merge, thus circumventing the so-called \emph{final-parsec problem} \citep{2003AIPC..686..201M}. The bottleneck to SMBH binary evolution \citep{1980Natur.287..307B} is believed to occur on the parsec scale, where intersecting-orbit stars have all been ejected but GWs are still not efficient enough to finally drive the inspiral. In principle, the relatively low-density gas-poor galaxy cores of  BCGs are the most exposed to SMBH binary stalling. It has been found that triaxial potentials might alleviate the problem by increasing the number of orbits that cross the binary's loss-cone, therefore providing a way to get rid of additional binary energy and angular momentum \citep{2004ApJ...606..788M}. However, a recent investigation by \cite{2014ApJ...785..163V} called this result into question by showing that triaxiality alone might not be enough. Nonetheless, in real mergers, other factors such as rotation, bar-like instabilities and an unrelaxed time evolving potential might significantly enhance the flux of stars into the loss cone \citep{2006ApJ...642L..21B}, and recent ab-initio N-body simulations of merging stellar bulges succeeded in driving the SMBH binary to final coalscence \citep{2011ApJ...732L..26P,2011ApJ...732...89K}. If some gas if present, this may provide additional help in hardening the binary (see, e.g., \citealt{2002ApJ...567L...9A,2005ApJ...630..152E,2007MNRAS.379..956D} for gas driven binaries), even though it has been also argued that gas might indeed be unable to absorb significant angular momentum from the binary if the gaseous-disk mass is limited by self-gravity and fragmentation \citep{2009MNRAS.398.1392L}. 

Secondly, we only update  SMBH masses and spins during merging events, thus neglecting any accretion mechanism. Giant ellipticals are gas-poor systems, generally unable to supply large amounts of material to feed the central SMBH. It is observationally well known that the accretion activity of the most massive black holes peaks at $z\approx2$ \citep[see, e.g.,][]{2007ApJ...654..731H}, rapidly declining at lower redshifts. This trend has been reproduced by state of the art theoretical models, which find that the most massive SMBHs at low $z$ grow primarily via mergers \citep{2007MNRAS.382.1394M,2011MNRAS.410...53F}, with little contribution from gas accretion.
The change of the SMBH spin magnitude due to accretion can also be safely neglected: momentum-conservation arguments \citep{1974ApJ...191..507T} imply that the spin magnitude is modified significantly only if the accreted mass is the order of the SMBH mass itself. This assumption is coherent with taking isotropic spin directions  neglecting further spin-alignment processes (see discussion in Sec.~\ref{bhmergermod}).

{Thirdly, we neglect any delay between galactic and SMBH binary mergers, thus assuming that they take place  simultaneously. In reality, binary formation and inspiral will postpone the SMBH merger even if the \textit{final-parsec problem} is solved efficiently. In dense stellar environments, if there is a continuous supply of stars interacting with the binary (technically, a full loss cone) SMBHs generally inspiral for $>3\times 10^7$yr before merging with each other \citep{2010ApJ...719..851S}, and similar timescales apply to gaseous environments \citep{2009MNRAS.396.1640D}. This delay will likely be longer for low density ellipticals \citep{2011ApJ...732...89K}; however, BCGs generally experience at most 2-3 major mergers since $z=1$, therefore delayed SMBH binary mergers could have a substantial impact on our results only if binaries typically survive for Gyrs (in which case, the distinction between delayed merger and stalling becomes blurry). We try here to critically assess the impact on delayed mergers on our results. We consider the longest merger timescales found in N-body simulations of merging galaxies performed by \cite{2011ApJ...732L..26P,2012ApJ...749..147K}. When scaled to massive ellipticals, the results of \cite{2012ApJ...749..147K} give coalescence times that can be as long as $\sim1$Gyr (see their Table 5), whereas \cite{2011ApJ...732L..26P} provide shorter timescales (see their Figure 4). We therefore count a posteriori the fraction of subsequent mergers separated by less than 1Gyr. This fractions turned out to be:}
\begin{itemize} 
{\item $\sim$0.2 in the "Fiducial" scenario; 
\item $\sim$0.3 in the "Optimistic" scenario;
\item $\sim$0.12 in the "Pessimistic" scenario; 
\item $\sim$0.25 in the "Pessimistic--Minor" scenario (however, in this latter case, also the number of mergers is larger).}
\end{itemize}
{We see that delayed mergers can produce triple interaction in 30\% of the cases at most (considering only the major merger statistics). When a triplet forms, either (i) a strong triple interaction occurs, causing the ejection of the lightest of the three SMBHs (and possibly accelerating the coalescence of the binary left behind), or (ii) a hierarchical system forms, possibly exciting Kozai resonances in the inner binary, again driving it to rapid coalescence. The outcome of the two processes is generally different, and the occurrence of one or the other depends on how far has the SMBH binary already gone into the hardening process, on how shallow has the galaxy core became, etc. We notice, however, that in case (i) the number of coalescences decreases at most proportionally with the fraction of triplets that forms, whereas in case (ii), the number of coalescences is basically unaffected, since each triplet formation leads to the coalescence of the binary that was already in place. Extensive numerical experiments performed by \cite{2007MNRAS.377..957H} showed that triple interactions generally lead to at least one binary coalescence (in 85\% of the cases), usually on a timescale shorter than 1~Gyr \citep[figure 8 in][]{2007MNRAS.377..957H}. Therefore, triple interactions might cause a fractional change of our ejection fractions of $0.3$ at most. In any case, it might be interesting to track consistently triplets in our simulations, and this point may be the subject of future improvements of our model. We also note that similar assumptions are also often made in more elaborate galaxy-evolution models (see e.g. \citealt{2012MNRAS.423.2533B} for a critical discussion).}

We are also neglecting the previous merger history of the BCGs. BCGs will generally reach $z=1$ after multiple merger events. The inspiral of a SMBH binary preceding a merger is expected to leave an imprint on the host galaxy in the form of a core scouring in the BCG center (especially if little nuclear star formation occurs). At each merger, the mass ejected in stars is of the order of $\sim 0.5 M$ (where $M$ is the total mass of the binary, \citealt{2006ApJ...648..976M}). The effect may be important after many merger generations and it leads to strong modification of the galactic potential in the core region. This effect is absent in our simplified model, but we note that the core properties are only important when estimating the SMBH return time in the Bounce model (Sec.~\ref{timebackmodel}). The fitting procedure developed here is built on the results obtained by \cite{2008ApJ...678..780G}, which in turn consider an elaborate galaxy model \citep{2005MNRAS.362..197T} where core depletion is taken into account.

%%%%%%%%%%%%%%%%%%%%%%%%%%%%%%%%%%%%%%%%%%
\section{Results and discussion}
\label{sec_results}

We combine different prescriptions for two main processes
\begin{itemize}
\item the return time: "Dynamical Friction" (DF) or "Bounce" (Sec.~\ref{timebackmodel});
\item the merger distribution: "Fiducial", "Optimistic" or "Pessimistic" (Sec.~\ref{mergermodel} ).
\end{itemize}
This results in a set of six models that we use as investigation playground: "Fiducial-DF", "Fiducial-Bounce", "Optimistic-DF", "Optimistic-Bounce", "Pessimistic-DF", "Pessimistic-Bounce". In each model, the evolution of the SMBH population is characterized by the following input parameters: 
\begin{itemize}
\item initial BCG occupation fraction $f_{\rm z=1}$; 
\item occupation fraction of the satellite galaxies $f_{s}$; 
\item initial SMBH spin magnitudes in the BCGs $\chi_{z=1}$;  
\item SMBH spin magnitudes in the satellites $\chi_s$. 
\end{itemize}
We discuss in the following the results of our simulations, separating the effect of each individual parameter. 
The main observables are:
\begin{itemize}
\item final BCG occupation fraction $f_{\rm z=0}$ (later splitted between those galaxies which underwent a SMBH replenishment $f^{\rm R}_{\rm z=0}$ and those which keep their original SMBH $f^{\rm NR}_{\rm z=0}$);
\item fraction of BCG that do not host a nuclear SMBH at $z=0$, simply defined by  $1-f_{\rm z=0}$;
\item distance from the BCG center (offset) of the ejected SMBH at the present time $r_{\rm z=0}$.
\end{itemize}

For any given set of parameters we simulate 1000 BCGs (with the exception of the runs presented in Figs.~\ref{replen} and {\ref{offset}} which contains $10000$ BCGs): typical Poisson counting errors on the final occupation fractions are therefore $\sim 3 \%$.
{Most of the results presented here (with the exception of Sec.~\ref{initial_f} where such issue is explicitly investigated) are computed assuming $f_{\rm z=1}=1$ as a simplifying assumption (cf. Sec.~\ref{sec_intro})}

%%%%%%%%%%%%%%%%%%%%%%%%%%%%%%%%%%%%%%%%%%%
\subsection{The impact of the host properties: cluster shape and BCG merger rates}
The six models described above are defined by distinct `environmental properties' which are not directly related to the SMBH population itself; namely the merger history of BCGs (determining the number of SMBH binary mergers) and the shape of the cluster potential (governing the typical return timescales of ejected SMBHs). We describe their impact on the results first (fixing $f_{z=1}=f_s=1$), turning to the properties of the SMBH population in the next subsection.
%%%%%%%%%%%%%%
\begin{figure*}
\includegraphics[width=0.9\textwidth]{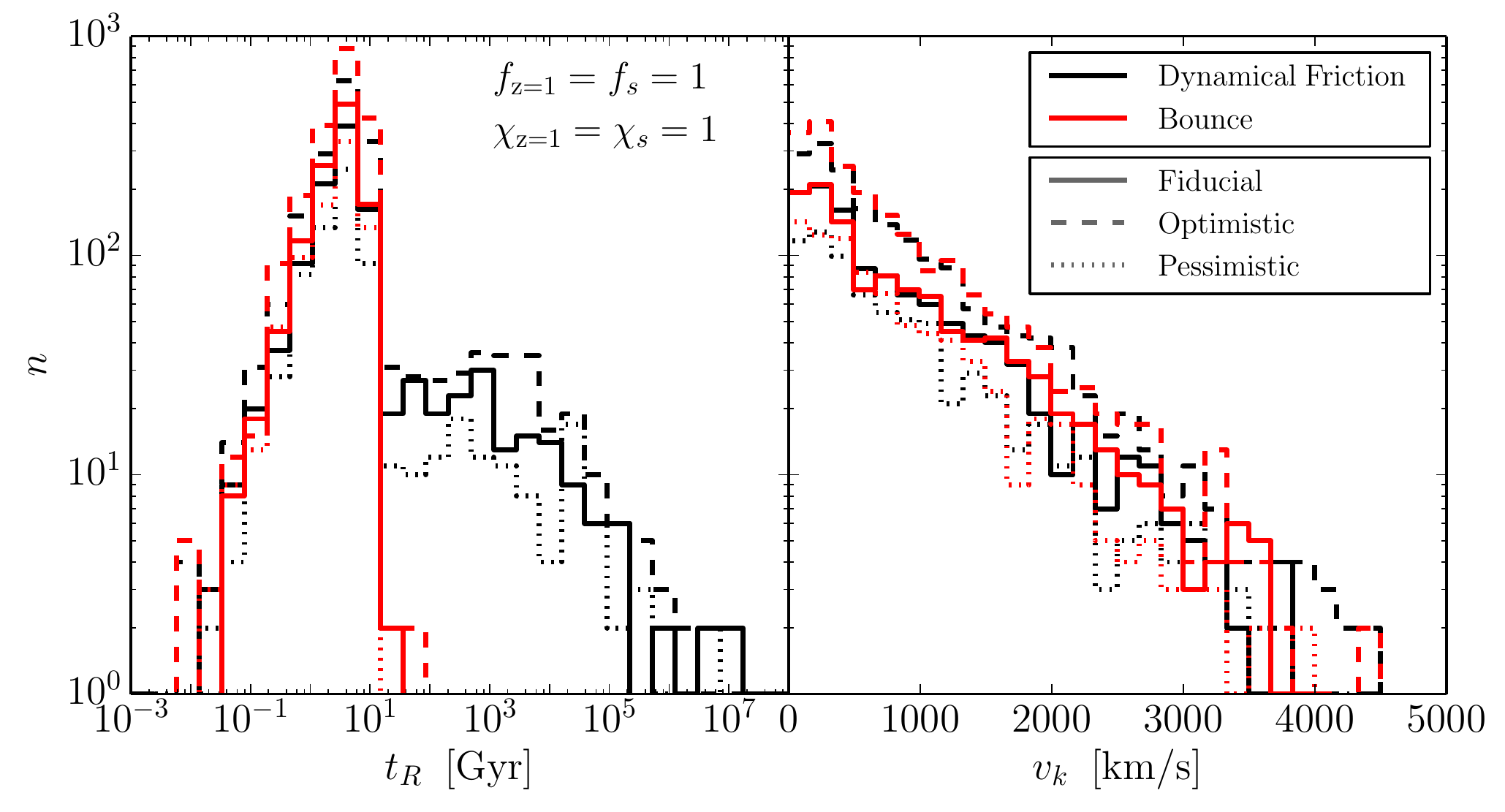}
\centering
\caption{(color online) Return time distribution $t_R$ (left) and recoil velocity distribution $v_k$ (right) of all  kicked SMBH in a 1000-events Montecarlo realization of our four fiducial models. {Red (black) curves are for the Bounce (DF) models, whereas solid, dashed and dotted curves correspond to the Fiducial, Optimistic and Pessimistic scenarios respectively, as labeled in figure.} The dotted vertical lines are the median values of all the distributions (which are not distinguishable on this scale). All distributions are computed assuming unity occupation fractions at $z=1$ and  $\chi_{z=1}=\chi_s=1$.} 
\label{tret}
\end{figure*}
%%%%%%%%%%%%%%
%%%%%%%%%%%%%%
\begin{figure*}
\begin{tabular}{cc}
\includegraphics[width=7.2cm]{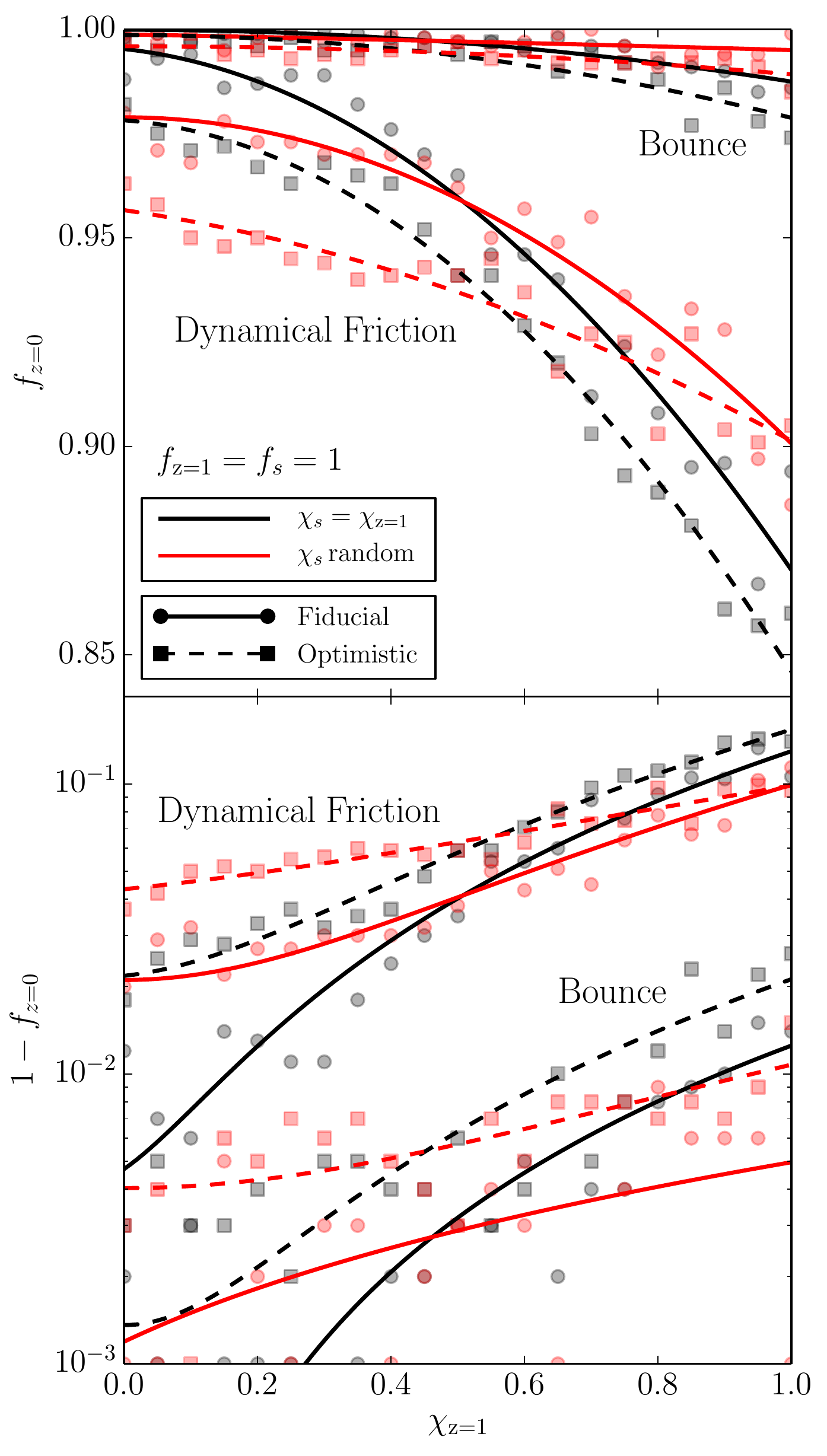}&
\includegraphics[width=7.2cm]{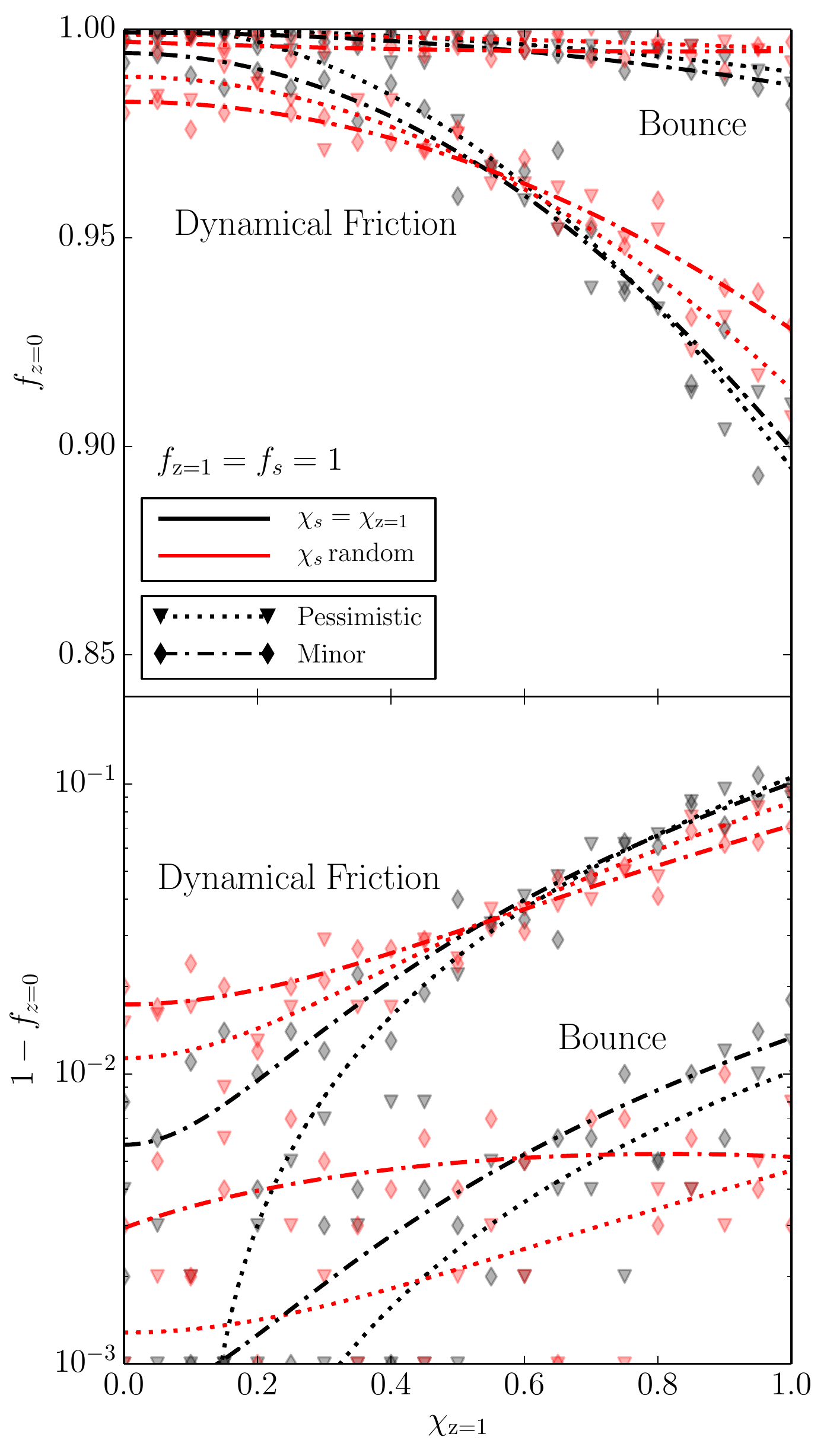}\\
\end{tabular}
\caption{{(color online) BCG occupation fractions. The left plot shows the "Fiducial" and the "Optimistic" models, whereas the right plots compares the "Pessimistic" and the "Pessimistic-Minor" models, to assess the impact of minor mergers. In each plot, the top panel shows the dependence of the $z=0$ occupation fraction $f_{\rm z=0}$ on the initial BCG spin magnitude $\chi_{z=1}$. To highlight the peculiarities of each individual model, the lower panel shows the corresponding depletion fraction $1-f_{\rm z=0}$, in logarithmic scale. Runs have been performed with two prescriptions on the spin magnitude of the satellite galaxy SMBHs $\chi_s$, taken either to be equal to the spins of the  BCG SMBHs   (black curves) or uniformly distributed in $[0,1]$ (red curves). A quadratic interpolation is presented in both cases. While final fractions as low as $\sim 0.85$ are detected in the DF scenario, only $f_{z=0}\sim 0.98$ can be achieved in spherically symmetric (Bounce) galaxies even for maximally spinning SMBHs.} 
}
\label{occfr_spin}
\end{figure*}
%%%%%%%%%%%%%%
%%%%%%%%%%%%%%%%%%%%%%%%%%%%%%%%%%%%%%%%%%%
\subsubsection{Bounce vs DF models}
The detailed shape of the cluster potential affects the trajectory of the recoiling SMBH. If all gravitational potentials were spherically symmetric, then SMBHs would always get back to the core of BCGs, and the Bounce model would provide a complete description of the dynamics. However cluster density profiles are often triaxial, unrelaxed, and `clumpy'. In a triaxial potential orbits do not conserve angular momentum, implying that the SMBH will miss the BCG core at subsequent passages; additionally, gravitational perturbations due to sub-halos and other galaxies can easily deflect the SMBH out of its initially radial orbit. The DF model is taken as an extreme (and admittedly unrealistic) case in which the SMBH returns on a circular orbit. Both the DF and the Bounce models are idealizations meant to bracket the range of possible outcomes. 
As shown in Fig. \ref{kick_df} for three selected systems, return timescales can easily exceed the Hubble time in the DF model. This is better seen in Fig. \ref{tret} where the distributions of recoil velocities $v_k$ and return times $t_R$ are computed along the evolution of the BCG population for our four default models. For all of them, the recoil distribution presents a high velocity tail extending to about $4000$ km/s$^{-1}$, with a median value of about $600$ km/s$^{-1}$. The difference between the Bounce and the DF models is clearly shown in the return time distribution. As expected, the rise of the distribution at $t_R<1$ Gyr (corresponding to small kick velocities) is similar because the bounce dynamics is basically equivalent to a DF process when the SMBH dot not leave the galaxy core. However, in the DF scenario, about 10\% of the SMBH are ejected outside the host BCG and interacts only with the low-density dark-matter background outside the galaxy, with resulting return times longer than $10$~Gyr (cf. the bump of the black distributions in the left panel of Fig.~\ref{tret}). As a result, BCG occupation fractions $f_{z=0}$ can be as low as 85\% in the case $\chi_{z=1}=\chi_s=1$, as reported in the upper panels of Fig.~\ref{occfr_spin}. Conversely, in the Bounce model, only few SMBHs do not make it back to the galaxy core following a kick, resulting in occupation fractions of 98\% or higher.
 The two models are best compared in terms of `depleted fraction', i.e. the fraction of BCGs that do not host a SMBH at $z=0$, which is simply $1-f_{z=0}$. This is shown in the lower panels of Fig.~\ref{occfr_spin}; it is clear that the DF model depletes BCGs of their central SMBH 10 times more efficiently than the Bounce model. 

%%%%%%%%%%%%%%%%%%%%%%%%%%%%%%%%%%%%%%%%%%%
\subsubsection{Fiducial, Optimistic and Pessimistic models}
{Conversely, the adopted merger rate does not have a strong impact on $f_{z=0}$, and the difference between Fiducial, Optimistic and Pessimistic models is only modest, being at most a factor of $\sim2$ in terms of depleted fractions, as shown in Fig.~\ref{occfr_spin}. For example, for  $\chi_{z=1}=\chi_s=1$, $1-f_{z=0}$ varies between 0.1 and 0.15. The impact of minor mergers is also small, as shown in the left panels of Fig.~\ref{occfr_spin}.}

\begin{figure*}
\includegraphics[width=0.95\textwidth]{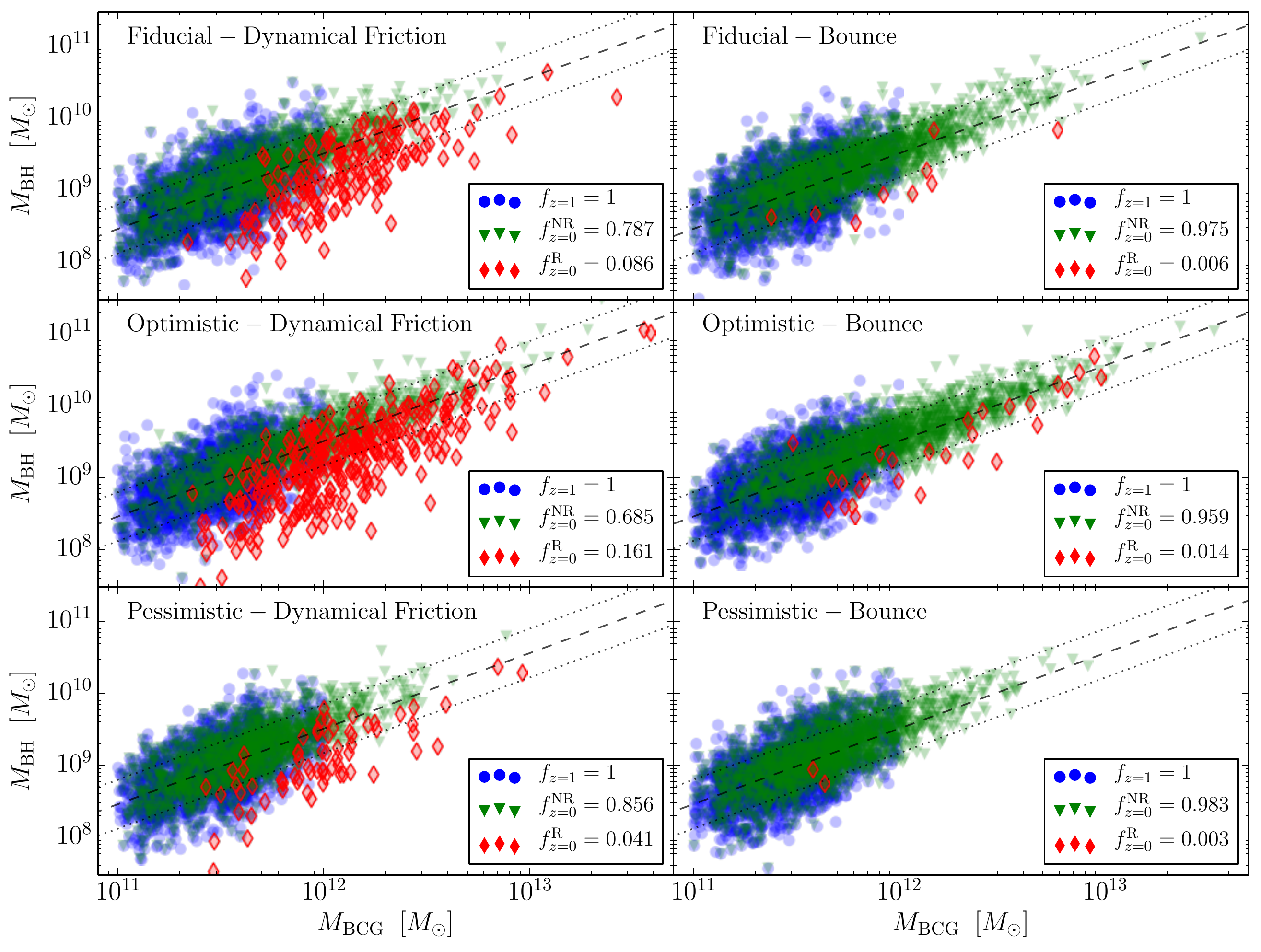}
\caption{(color online) Deviations from the SMBH/host relation in replenished galaxies, and final occupation fractions. We show the distributions of the SMBH mass $M_{BH}$ and the galaxy mass $M_{BCG}$ in our six different models, assuming $f_{\rm z=1}=f_s=1$ and $ \chi_{\rm z=1}=\chi_s=1$. Dashed and dotted lines show the average and the standard deviation of the initial correlation (\ref{mbhmgal}). Blue circles shows the initial $z=1$ sample. We track those system where a replenishment occurred (R, red diamonds) and those which just underwent a plain evolution to $z=0$ (NR, green triangles). While the evolved NR sample still lies on the $z=1$ correlation but with lower scatter, replenished galaxies clearly exhibit deviations towards lower $M_{\rm BH}$ values. {Occupation fractions for each sample are reported in the legends and are computed considering 10000 initial BCGs; points are shown for only 2000 initial BCGs to avoid cluttering.} 
}
\label{replen}
\end{figure*}

Although apparently counter intuitive, this result is in fact expected because a higher BCG merger rate implies also a higher probability of multiple mergers. While it is true that each SMBH has a larger chance to be kicked out of its host, it is also true that there is a higher probability that it is replaced by another (possibly undermassive) SMBH brought in by a subsequent merger. Enhanced ejections and replenishments nearly cancel out making $f_{z=0}$ only weakly dependent on the details of the merger history. This is illustrated in Fig. \ref{replen}, where the extreme case $\chi_{z=1}=\chi_s=1$ is considered. In the Fiducial-DF model, 87\% of the BCGs host a SMBH at $z=0$ ($f_{z=0}=0.87$); however, only 79\% of them retained their original $z=1$ SMBH, while $\sim$9\% are depleted of their original SMBH and `replenished' in a subsequent merger with a satellite galaxy hosting a SMBH. In the Optimistic-DF model those percentages become 69\% and 16\%, respectively: more SMBH are ejected (only 69\% of original SMBHs retained), but a larger fraction of BCGs is replenished (16\%) by virtue of the higher merger rate (causing a higher probability of multiple mergers). {The opposite behavior is detected when the Pessimistic-DF scenario is considered. The balance is almost perfect in the Bounce models (also shown in Fig. \ref{replen}). All three scenarios show $f_{z=0}\simeq 0.98$, but  the probability of replenishment increases from the Pessimistic to the Fiducial and Optimistic models following a larger number of SMBH ejections. }

{As expected, the SMBH-mass distributions are different for replenished and non-replenished galaxies. Non-replenished galaxies reflect the injected correlation law (\ref{mbhmgal}) with lower scattering at $z=0$, while the replenished samples  tend to host undermassive SMBHs which have grown within smaller satellite galaxies in the cluster.}
%%%%%%%%%%%%%%
%%%%%%%%%%%%%%

%%%%%%%%%%%%%%%%%%%%%%%%%%%%%%%%%%%%%%%%%%%
\subsection{The impact of the SMBH properties: spin magnitude and initial occupation fraction}
Having explored the impact of the physics governing the evolution of the SMBH environment, we turn now to a  description of  the effect of the parameters related to the SMBH population itself; in particular SMBH spins and initial occupation fraction.   

%%%%%%%%%%%%%%%%%%%%%%%%%%%%%%%%%%%%%%%%%%%
\subsubsection{Spin magnitude}
The magnitude of the SMBH spin vectors in BCGs is essentially unknown, since most of the direct measurements from K$\alpha$ iron lines involve local Seyfert galaxies \citep{2013mams.book.....B,2013SSRv..tmp...81R} and it is difficult do derive clear constrains through indirect arguments related to jet production, AGN spectra energy distribution fitting, or the evolution of the SMBH accretion efficiency with mass and redshift \citep[see, e.g.,][]{2009ApJ...697L.141W,2010ApJ...718..231S,2014MNRAS.438..672N}. However, we know that spins are crucial in the physics of gravitational recoils, because highly spinning SMBHs are likely to experience stronger recoils [see Eq.~(\ref{vparallel})]. We therefore need to investigate the SMBH parameter space carefully, to cover the full range of possibilities predicted by our models. For each of our four models, we initialize $\chi_{z=1}$ 
at a fixed value, running between zero and one. {As stated in Sec.~\ref{bhmergermod}, the spin orientations are assumed to be isotropic.} For each case, we consider two different $\chi_s$ distributions: i) $\chi_s=\chi_{z=1}$ in each individual merger, and ii) $\chi_s$ random in the range $[0,1]$. As shown in the upper panels of Fig.~\ref{occfr_spin}, $f_{z=0}$ is always a decreasing function of $\chi_{z=1}$, and is fairly well described by a quadratic function. Trends are best seen in the lower panels of Fig.~\ref{occfr_spin}, where we plot the depleted BCG fraction $1-f_{z=0}$. 
In terms of the depleted fraction, spins have an order of magnitude impact on the results. In the DF model, only $\approx$1-4\% of the BCGs are depleted at $z=0$ (i.e., $1-f_{z=0}=0.01-0.04$) for $\chi_{z=1}=0$, whereas up to $\approx$10-15\% of the BCGs lost their SMBH at $z=0$ (i.e., $1-f_{z=0}=0.1-0.15$) for $\chi_{z=1}=1$. Similar trends hold for the Bounce model, but in that case only $\approx$0.1\% to $\approx$2\% of the SMBHs are lost at $z=0$. It is interesting to notice that even for $\chi_{z=1}=0$, we get $0.01<1-f_{z=0}<0.04$ in the DF models. This is, again, because of multiple mergers: a Schwarzschild SMBH can acquire a spin $\chi\approx 0.5-0.6$ in a single merger event [see Eq.~(\ref{finalspin})], which significantly enhances the probability to experience a superkick if a subsequent merger occurs. The different $\chi_s$ prescriptions [case i) and ii) above] show the same qualitative feature. The fits to the depleted fractions (lower panel of Fig.~\ref{occfr_spin}) intersect around $\chi_{z=1}=0.5$ as expected: for lower values, the average $\chi_s$ in case ii) is larger, resulting in more superkicks and more SMBH ejections, while the opposite is true in case i).        
  
%%%%%%%%%%%%%%%%%%%%%%%%%%%%%%%%%%%%%%%%%%%
\subsubsection{Initial BCG occupation fraction}
\label{initial_f}
All theoretical models developed to reproduce the SMBH cosmic evolution (including present number density, and quasar luminosity function 
up to high redshift) require an amount of SMBHs that guarantees an occupation fraction $f=1$ for massive galaxies \citep[see, e.g.][]{2007MNRAS.382.1394M,2011ApJ...742...13B,2011MNRAS.413..101G,2014arXiv1402.0888K}, pending, of course, the occurrence of superkicks. There is always the possibility that a superkick occurs at $z>1$, even though galaxies at higher redshift are generally richer of cold gas, which will likely promote SMBH spin alignment during mergers \citep{2007ApJ...661L.147B,2010MNRAS.402..682D}, ultimately suppressing superkicks \citep{2010ApJ...715.1006K}. Nonetheless, this might introduce some uncertainty on $f_{z=1}$ and, although we do not expect it to be far from unity, we study the sensitivity of our models to this parameter for completeness.

\begin{figure}
\includegraphics[width=\columnwidth]{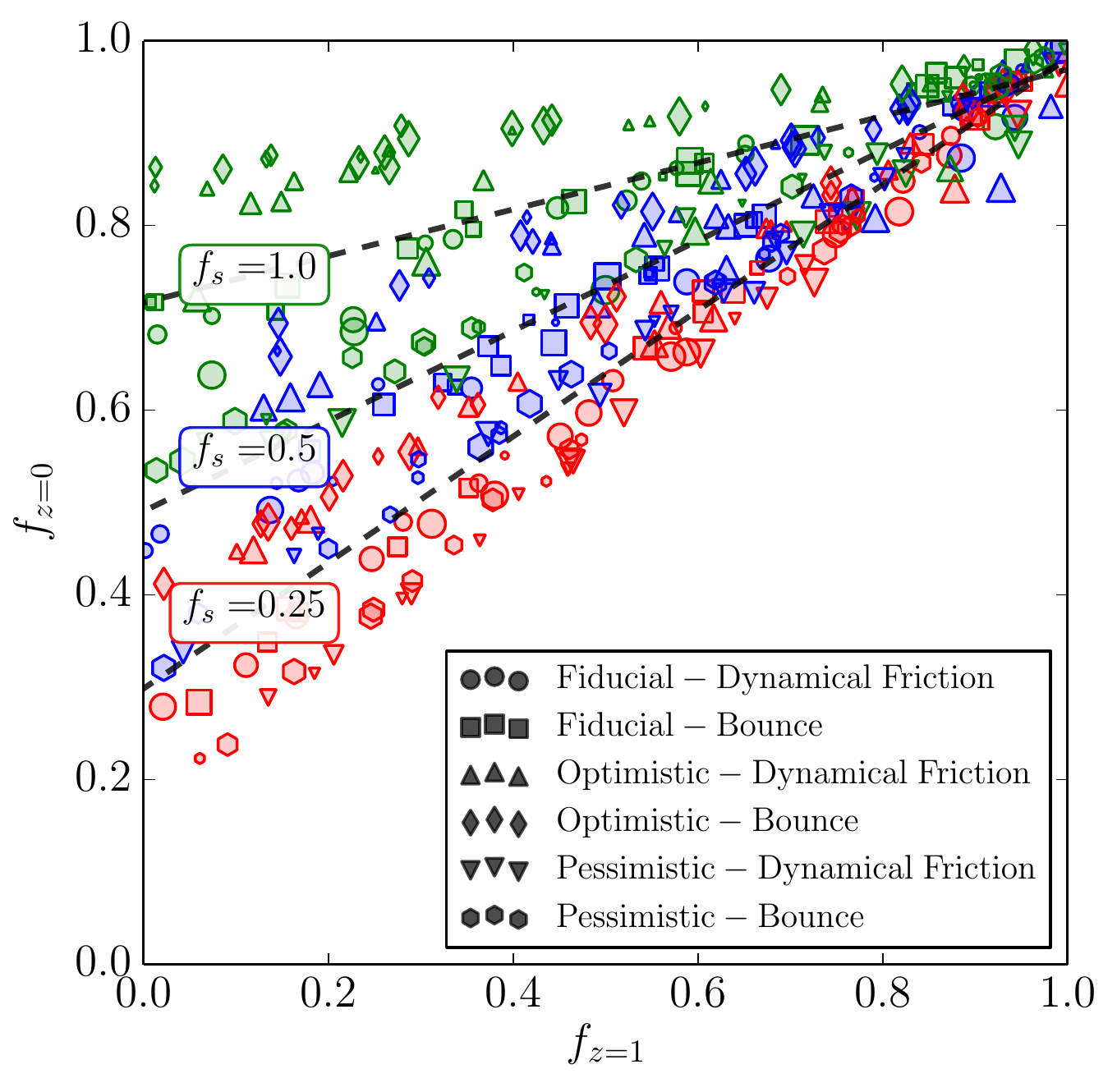}
\caption{(color online) Dependence of the final BCG occupation fraction  $(f_{z=0})$ on the initial occupation fraction of BCGs $(f_{z=1})$ and satellites ($f_{s}$). Each point represents a simulation of 1000 clusters where both the initial BCG and the satellites galaxies have the same initial spin  
$\chi_{\rm z=1}=\chi_s$ (indicated with symbol size, where small symbols stand for slowly rotating SMBHs and large symbols for high spins) and model prescriptions (indicated with different symbol shapes, as detailed in the legend).
Each sample 
(clustered along the dashed lines, with different colors)
 is computed with a different value of $f_s$. As confirmed analytically, the final BCG occupation fraction scales linearly with the initial occupation fraction; slopes and intercept are mainly determined by $f_s$. 
}
\label{occfr_fs}
\end{figure}

Fig.~\ref{occfr_fs} shows 
$f_{z=0}$ as a function of 
$f_{z=1}$, for 240 different merger trees. The main evidence is that $f_{z=1}$ scales linearly  with $f_{z=0}$.
The slopes and the intercept of the linear relation mostly depend on the occupation fraction of the satellite galaxies $f_{\rm s}$, i.e. on how many SMBHs are injected in the simulations between $z=0$ and $z=1$.
%%%%%%%%%%%%%%
%%%%%%%%%%%%%%
The linear relationship between $f_{z=1}$ and $f_{z=0}$ can  be easily understood using a simple analytic model (built on the line of \citealt{2007ApJ...667L.133S}). The probability  $f_i$ of a BCG to have a SMBH at the $i$-th merger generation consists in the sum of {(i)} the probability that only the BCG had a SMBH at the previous generation $f_{i-1}(1-f_s)$, {(ii)} the probability that only the satellite had a SMBH $f_s(1-f_{i-1})$ and {(iii)} the probability that there has been a merger but the SMBH has not been ejected $f_s f_{i-1} (1-P_{\rm ej})$ (where $P_{\rm ej}$ is the ejection probability).
This yields
\begin{align}
f_{i}=f_{s}+f_{i-1}-f_s f_{i-1}(1-P_{\rm ej})\,.
\end{align}
Using the convergence limit  $f_\infty = 1/(1+P_{\rm ej})$, and fixing $f_{z=1}$ as initial condition, we can write down the previous expression as a geometric progression 
\begin{align}
f_{i}-f_{\infty}=(f_{z=1} -f_\infty)\left(1-\frac{f_s}{f_\infty}\right)^i \,.
\end{align}
With the further (strongly idealized) assumption that $P_{\rm ej}$ is constant over different merger generations, we can estimate the final occupation fraction in our samples to be
\begin{align}
f_{z=0}-f_{\infty}=(f_{z=1} -f_\infty)\sum_{j=0} \epsilon_j \left(1-\frac{f_s}{f_\infty}\right)^j \,,
\end{align}
where $\epsilon_j$ is the fraction of BCG in which $j$ mergers occur between $z=1$ and $z=0$.
The above expression confirm the main trends observed in the simulations presented in Fig.~\ref{occfr_fs}, namely the linear relationship between  $f_{z=0}$ and $f_{z=1}$, with slope and intersect mainly depending on $f_s$. The initial occupations $f_{z=1}$ and $f_s$ are physically determined by cosmic history at early times ($z>1$), whose modeling is outside the scope of the present paper. However, as discussed before, we expect any deviation of $f_{z=1}$ from unity to be also related to the occurrence of superkicks.

%%%%%%%%%%%%%%%%%%%%%%%%%%%%%%%%%%%%%%%%%%%
%%%%%%%%%%%%%%
\begin{figure*}
\includegraphics[width=\textwidth]{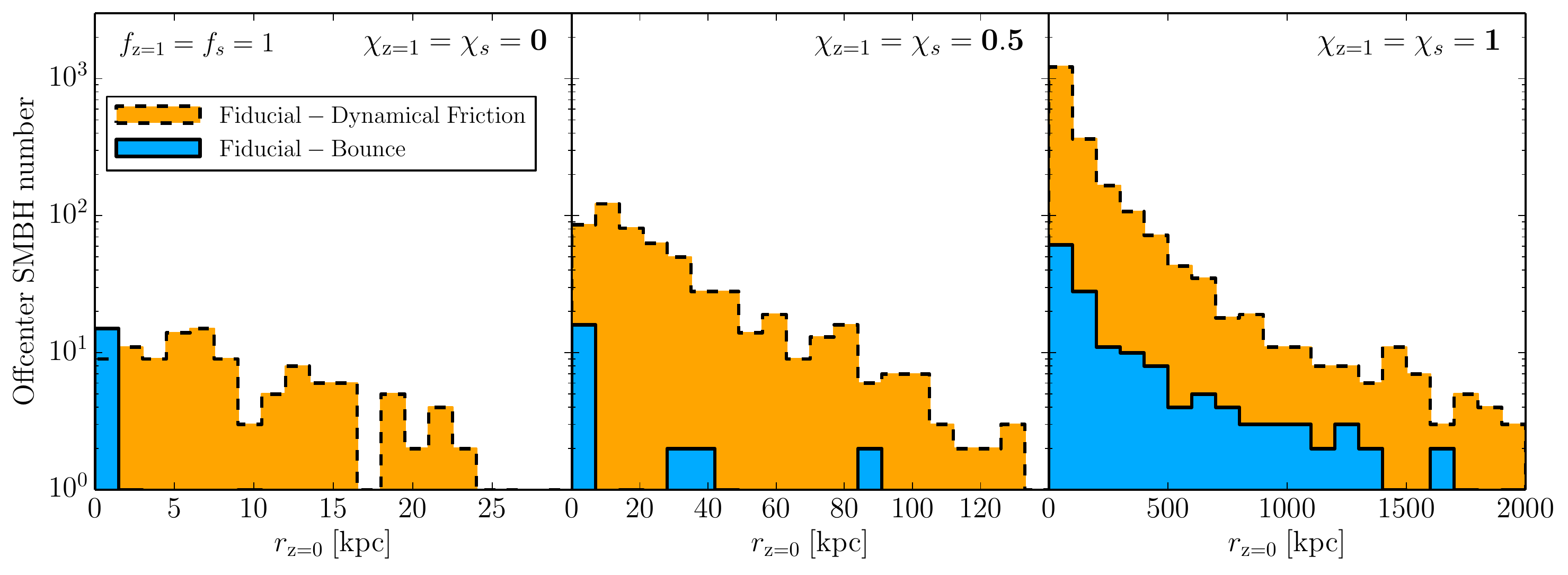}
\caption{(color online) Number of wandering, offcenter, SMBHs as a function of their present distance from the galactic center (offset) $r_{\rm z=0}$. The DF models (suited for non-spherical potentials) present at least a factor $\sim 10$ more wandering SMBHs than the Bounce models. SMBH detection through off nuclear quasar signatures or compact stellar systems may therefore distinguish between the two scenarios. Each run presented in this figure contains $10000$ BCGs, which sets the absolute scale of the SMBH number;  three different spin-magnitude values $\chi_{z=1}=\chi_s=0$ (left panel), $0.5$ (middle panel) and $1$ (right panel) are considered; initial occupation fractions are fixed at $f_{z=1}=f_s=1$.
}
\label{offset}
\end{figure*}
%%%%%%%%%%%%%%
\subsection{Discussion}
Our results show that superkicks likely have very interesting and potentially observable astrophysical consequences, most notably, a decrease of the SMBH occupation fraction in BCGs down to 0.9 or lower, under specific assumptions. At the time of writing, secure SMBH mass measurements have been performed in about 10 BCGs \citep{2012ApJ...756..179M}, an insufficient number to empirically constrain the models presented here. As described in the introduction, future 30m class telescopes like  ELT and TMT can easily boost those figures by a factor of 10 or more. {With $\mathcal{O}$(100) SMBH mass measurements, significant deviations from $f_{z=0}=1$ can be measured, making possible to directly test our superkick models, and possibly providing insights on the BCG SMBH spin distribution.} As shown in the previous section, $f_{z=0}$ strongly depends on both spin magnitudes and the detailed shape of the cluster potentials. The effect of the two ingredients is somewhat degenerate, since both high spins and non spherical potentials tend to reduce the occupation fraction. The degeneracy is, however, only partial. For example, $f_{z=0}<0.9$ is possible only if cluster potentials are extremely non-spherical {\it and} typical spins are higher than 0.8. A measurement of such low BCG occupation fraction will therefore provide valuable information on both the dynamics of the kicked SMBHs and their spins. Conversely, an occupation fraction of, say, 0.98 can be due to a combination of extremely low spins and non spherical potentials {\it or} very high spins and almost spherical potentials, as demonstrated in the lower panel of Fig. \ref{occfr_spin}. In this case, degeneracy might be broken via independent measurements of the cluster mass distribution derived, for example, by lensing. Those allow us to reconstruct the shape of the cluster potential, thus providing an estimate of how likely/unlikely it is for an ejected SMBH to return on a radial orbit.

{We note that the Bounce and DF prescriptions have been taken as extreme cases of a continuum range of possibilities. Since those prescriptions have a strong impact on the results, we can try to assess which of the two might be closer to reality on the basis of qualitative theoretical arguments. In the Bounce model, subsequent passages of the SMBH across the BCG core are crucial in damping the radial oscillations, critically shortening the return time. As a matter of fact, the clumpyness of a typical galaxy cluster mass distribution might easily cause a SMBH kicked to a few hundred kpc to miss a galaxy core which is smaller than 10~kpc across  \citep{2007ApJ...662..808L}. A simple estimate of the deviation from the radial path can be done by considering close encounters between the kicked SMBH and other cluster galaxies at apoastron. Consider a SMBH ejected at $r\approx100$~kpc in a typical cluster of $M_{\rm DM}=5\times10^{14}\msun$. The typical time it spends close to apoastron is $\delta{t}\approx0.1$~Gyr. The gravitational pull of a galaxy with mass $M$ at a distance $d$ from the SMBH, will cause a velocity change} 
\begin{equation}
\delta{v}\approx\frac{GM}{d^2}\delta{t}\approx50 \left(\frac{M}{10^{10}\msun}\right) \left(\frac{d}{10^{4}pc}\right)^{-2}~{\rm km\,\, s}^{-1}.
\end{equation}
{In a galaxy cluster like Coma, the galaxy density at 100kpc from the center is few$\times10^3$ galaxy Mpc$^{-3}$ \citep{2014MNRAS.441.3083W}, implying that the presence of at least one perturber at 
\mbox{$d<10$~kpc}
 is guaranteed. Considering a circular velocity of $v_c\approx10^{3}$~km s$^{-1}$, it is therefore very likely that SMBHs kicked at $r>100$~kpc will acquire a tangential velocity component $\approx0.1v_c$ because of interactions with nearby cluster galaxies (and clumpyness of the DM halo). We performed a simple test of the DF return timescales for non circular orbits by numerically integrating the DF equations in a Hernquist$+$NFW potential. We placed the sinking SMBH at a distance $R$ from the center, and we gave it an initial velocity $v=v_c(R)$ and $v=0.1v_c(R)$. The first case corresponds to a circular orbit, while the second implies a nominal eccentricity of $e\approx0.98$ (if the orbit was Keplerian). Despite an almost radial orbit, the return timescale in the latter case was only approximately 5 times shorter. We tested that reducing  $t_{\rm DF}$ in Eq. (\ref{DFtimescale}) by a factor 5 may cause a maximum variation of $\sim 0.07$ on $f_{z=0}$ in the extreme case $\chi_s=\chi_{z=1}=1$, which still implies $f_{z=0}\approx0.9$. This suggests that small deviations from a perfectly radial orbit result in return timescales just a factor of a few smaller than our DF computation, but two orders of magnitude longer than the Bounce model prediction, which is therefore relevant only for almost spherical potentials.} 
We conclude that the DF scenario provides a better approximation for the return timescales in realistic potentials implying interesting observational prospects. As shown in Fig. \ref{occfr_spin}, a DF-like dynamics results in $f_{z=0}<0.99$ for basically any choice of other relevant parameters, and the superkick effect should be detected  with a sample of $\mathcal{O}$(100) SMBH mass measurements.

Besides the lower BCG occupation fraction, another interesting phenomenon is BCG replenishment. We saw in the previous section that depleted BCGs can be replenished in a subsequent merger with another SMBH carried by the satellite galaxy. In this case, the new SMBH will most likely be undermassive with respect to the BCG mass. This is  shown by the red diamonds in Fig. \ref{replen}, which lie $\approx0.3$~dex below the SMBH-bulge relation defined by the green triangles. However, the net effect of replenishment is just to produce a slightly lower normalization and larger scatter in the SMBH-bulge relation, which would be hard to identify observationally.  

The implications of superkicks on the BCG occupation fraction are directly mirrored in the presence of a complementary population of wandering SMBHs. In fact, as already noted, full ejections from galaxy clusters are extremely unlikely because of the high escape speeds. As a natural consequence, some recoiled SMBHs are still sinking back to the BCG center today, and can potentially be detected as off center objects, adding evidence to the superkick scenario. Because of the longer return timescales, offcenter SMBHs are expected to be at least 10 time more likely in the DF than in the Bounce models. 
{The offset distribution is shown in Fig.~\ref{offset} for three values of the the spin magnitudes $\chi_{z=1}=\chi_s=0,0.5,1$, assuming the "Fiducial" model (other models, not shown, yield similar results)}. The absolute number of recoiling SMBHs in each panel is directly related to the average kick velocity imparted after SMBH mergers, which reflects the average spin magnitude. Distributions are generally monotonically decreasing functions of the offset $r_{z=0}$, meaning that many of these wandering SMBHs are concentrated in a few central kpc. However, in the maximally spinning case (right panel) about 50\% of the ejected SMBHs are located well outside the central BCG, with an offset between 100~kpc and 1~Mpc. Moreover, a tail extending to few Mpc is present, implying that a few SMBHs might even lurk in the outskirts of galaxy clusters. For this favorable configuration, we predict that between 0.5\% and 5\% of massive galaxy clusters should host a wondering BCG SMBH with an offset of a few hundred kpc from the cluster center. {The situation is less promising for lower spin values, even though in the intermediate case (central panel) for the DF model, about 1\% of the BCGs might host  a SMBHs lurking at few tens of kpc from their centers.}

Several observational signatures of recoiling SMBHs have been proposed in the literature, ranging from off center AGNs \citep{2011MNRAS.412.2154B} and tidal disruptions \citep{2008ApJ...683L..21K,2012ApJ...748...65L}, to intracluster ultracompact stellar systems \citep{2009ApJ...699.1690M}. All of them rely on the fact that the recoiling SMBH is carrying with it a significant amount of nuclear gas and stars, which is not likely in our case. Firstly, BCGs are mostly gas poor systems with shallow stellar cores; little cold gas should be available in the surrounding of the merger remnant, disfavoring off--nuclear AGN activity. Secondly, the SMBH can carry away only material that is orbiting around it with a velocity greater than the kick velocity 
$v_k$.
Ejections to a few hundred kpc require  
$v_{\rm k}>1500$ km s$^{-1} \gg \sigma$, implying that the mass in stars and gas that can be carried away is likely $<1\%$ of the SMBH mass. Lastly, because of their high mass, those SMBHs will simply swallow stars without tidally disrupting them, inhibiting the tidal disruption channel as a possible observational signature. The only possibility seems therefore to be the challenging detection of a faint ultracompact cluster with extremely high velocity dispersion, which might be feasible in nearby galaxy clusters as discussed by \cite{2009ApJ...699.1690M}. Alternatively, also `naked' SMBHs still interact with the diffused hot intracluster gas. This can produce X-ray emission potentially observable at nearby galaxy cluster distances (see \citealt{2009MNRAS.394..633D} for details). 

%%%%%%%%%%%%%%%%%%%%%%%%%%%%%%%%%%%%%%%%%%%%%%%%%%%%%%%%%%%%%%

\section{Summary and conclusions} \label{sec_concl}
In this paper we investigated the consequences of superkicks for the population of the most massive SMBHs in the Universe residing in BCGs. 
The choice of BCGs as study targets follows from a number of theoretical and observational arguments: i) compared to other types of galaxies, BCGs have the richest merger history, especially at low redshift, ii) future 30m-scale telescopes will have the resolution to easily reveal SMBHs in hundreds of BCGs up  to $z\approx0.2$, iii) theoretically, BCGs are expected to have unit SMBH occupation fraction, and even a single depleted system would be the smoking gun of superkick occurrence in nature. We demonstrate that, under plausible astrophysical assumptions, SMBHs can be ejected from BCG cores, potentially resulting in an occupation fraction substantially lower than one in the local Universe (say, $z<0.1$). 

Starting from the observational fact that BCGs have doubled their mass since $z=1$ -- and that this mass growth is consistent with their merger activity as inferred from galaxy pair counts, and as found in simulations of galaxy formation -- we have constructed a simple semianalytical model to track their evolution to the present time. Our model reconstructs the dynamics of each single major merger, including a self-consistent computation of the gravitational recoil and of the return time of the kicked SMBHs. We considered six classes of models combining two BCG major merger history models ("Fiducial", "Optimistic" and "Pessimistic", covering the range consistent with observations and simulations) and two specific prescriptions for the return times (``Bounce" and ``DF"). Minor merger rates were also available for the "Pessimistic" scenario, we investigated their impact by including them in the "Pessimistic-Minor" model. Since the magnitude of the spins of SMBHs in BCGs is basically unknown, for each model we considered a range of spin distributions for the SMBHs residing in the BCGs, $\chi$, and in the merging satellites, $\chi_s$. We ran several sets of simulations varying all the relevant parameters, we studied their impact on the final BCG occupation fraction $f_{z=0}$, and we investigated possible observational consequences.

Our main results can be summarized as follows:
\begin{enumerate}
\item superkicks can efficiently deplete BCGs of their central SMBHs. The occupation fraction at $z=0$ can be as low as $f_{z=0}=0.85$ for the most favorable scenarios;   
\item $f_{z=0}$ is quite insensitive to the BCG merger history, so long as those experience at least $\approx1$ merger since $z=1$;
\item {only small quantitative differences were found when comparing the "Pessimistic" and the "Pessimistic-Minor" models, implying that the poorly constrained distribution of minor mergers is not a significant caveat to our findings;}
\item $f_{z=0}$ is very sensitive to the dynamics of the ejected SMBHs in the galaxy cluster potential well. The fraction of depleted BCGs (i.e. $1-f_{z=0}$) is of the order of 0.01 only for the Bounce models, but it is typically 0.05-0.1 for the DF models;
\item the intial value of the SMBH spins has an order of magnitude influence on the depleted BCG fraction. In the DF models, this varies from $\approx0.02$ for non spinning SMBHs, up to $\approx0.15$ for maximally spinning SMBHs;
\item we predict that few a percent of the galaxy clusters host an offset BCG SMBH inspiralling at a few hundred kpc from the dynamical center, although they might be extremely difficult to detect;
\item for a large variety of physically plausible scenarios, we predict $f_{z=0}<0.99$, that can be directly tested with measurements of SMBHs in the center of $\mathcal{O}$(100) BCGs with future 30m telescopes. 
\end{enumerate}

As detailed in Sec. \ref{sec2.5}, we made a number of simplifying assumptions in our calculation. In particular we neglected any possible mass and spin evolution due to gas accretion, and we assumed SMBH binaries always merge following galaxy mergers (i.e., we by-passed the \emph{final parsec problem}). Moreover, we assumed random spin orientations when computing kick velocities. We showed that all these assumptions are well justified at least for the majority of mergers involving BCGs, but refinement of some of them might be considered for future work. 

Although current statistics of SMBH mass measurements in BCGs is insufficient to empirically constrain the models presented here, prospects look promising for the next generation of 30m-class optical telescopes. Any measurement of a BCG occupation fraction lower than unity will provide observational evidence for the occurrence of superkicks in nature, bringing the extreme dynamical effects of strong-field general relativity to the realm of observational astronomy.

%%%%%%%%%%%%%%%%%%%%%%%%%%%%%%%%%%%%%%%%%%
\section*{acknowledgments}
We thank Emanuele Berti, Ulrich Sperhake, Giovanni Rosotti, Enrico Barausse, Tod Lauer, Marc Postman and Christopher Reynolds for helpful discussions.
DG is supported by the UK Science and Technology Facility Council and the Isaac Newton Studentship of the University of Cambridge; partial support is also acknowledged  from 
the FP7-PEOPLE-2011-CIG Grant No. 293412 ``CBHEO";
the FP7-PEOPLE-2011-IRSES Grant No. 295189 ``NRHEP'';
the STFC GR Roller Grant No. ST/L000636/1;
the Cosmos system, part of DiRAC, funded by STFC and BIS under Grant Nos. ST/K00333X/1 and ST/J005673/1;
the NSF XSEDE Grant No. PHY-090003;
the CESGA-ICTS Grant No. 249  
and the NSF CAREER Grant No. PHY-1055103. 
AS is supported by the DLR (Deutsches Zentrum fur Luft- und Raumfahrt) through the DFG grant SFB/TR 7 Gravitational Wave Astronomy. DG finally thanks the kind hospitality received at the AEI where this work was conceived.  Most figures have been generated using the \texttt{Python}-based
\texttt{matplotlib} package \citep{2007CSE.....9...90H}.

{\footnotesize
\bibliographystyle{mn2e} 
\bibliography{bcg} }

\end{document}